\documentclass[journal,twoside,web]{ieeecolor}
\usepackage{generic}
\usepackage{graphicx,algorithmicx,algorithm,algpseudocode,verbatim,mathtools,enumerate,bm,hyperref,dsfont,tikz,pgfplots,cleveref,subcaption,nicefrac,booktabs}
\usepackage{cite}
\usepackage{amsmath,amssymb,amsfonts}
\usepackage{graphicx}
\usepackage{hyperref}
\usepackage{balance} 
\usepackage[acronym]{glossaries}
\usepackage{textcomp}
\hypersetup{hidelinks}
\crefformat{equation}{(#2#1#3)}
\usepackage[fancythm,fancybb]{jphmacros2e} 

\def\BibTeX{{\rm B\kern-.05em{\sc i\kern-.025em b}\kern-.08em
    T\kern-.1667em\lower.7ex\hbox{E}\kern-.125emX}}
\usepackage{xcolor}

\newenvironment{customproof}[1][Proof]{%
  \par\noindent\textbf{#1.}\quad%
}{\hfill$\frQED$\\}

\newacronym{adsb}{ADS-B}{Automatic Dependent Surveillance-Broadcast}
\newacronym{kkt}{KKT}{Karush-Kuhn-Tucker}

\newcommand{\mnorm}[1]{{\left\vert\kern-0.25ex\left\vert\kern-0.25ex\left\vert #1 
    \right\vert\kern-0.25ex\right\vert\kern-0.25ex\right\vert}}

\newcommand{\ie}{{\it i.e.}}

\newcommand{\fix}[1]{{#1}}
\newcommand{\fixx}[1]{{#1}}

\newcommand{\intSet}{\mathbb{Z}}

\newcommand{\actionPool}{\mathbb{X}}
\newcommand{\xSet}{\mathbf{x}}
\newcommand{\payList}{\mathcal{P}}
\newcommand{\offerList}{\mathcal{O}}
\newcommand{\profitList}{\mathcal{J}}
\newcommand{\costList}{\mathcal{C}}
\newcommand{\separationDist}{D}
\newcommand{\cSet}{\mathbf{c}}

\newcommand{\valuation}{\mathbf{b}}

\newcommand{\cycle}{\sigma}
\newcommand{\decFactor}{\gamma}

\newcommand{\smax}{S_i^{\mathrm{max}}}
\newcommand{\smin}{S_i^{\mathrm{min}}}

\begin{document}
\bstctlcite{IEEEexample:BSTcontrol}
\title{Noncooperative \fixx{Coordination} via a \\ Trading-based Auction}
\author{Jaehan Im, \IEEEmembership{Graduate Student Member, IEEE}, Filippos Fotiadis, \IEEEmembership{Member, IEEE}, Daniel Delahaye, \IEEEmembership{Member, IEEE}, Ufuk Topcu, \IEEEmembership{Fellow, IEEE}, David Fridovich-Keil, \IEEEmembership{Senior Member, IEEE}
\thanks{J. Im, U. Topcu, and D. Fridovich-Keil are with the Department of Aerospace Engineering and Engineering Mechanics, The University of Texas at Austin, TX, 78712, USA (emails: jaehan.im@utexas.edu,\, utopcu@utexas.edu,\, dfk@utexas.edu). F. Fotiadis is with the Oden Institute for Computational Engineering and Sciences, The University of Texas at Austin, TX, 78712, USA (email:  ffotiadis@utexas.edu). D. Delahaye is with Ecole Nationale de l'Aviation Civile (ENAC), 31055 Toulouse, France (email: daniel@recherche.enac.fr).}
\thanks{This work was supported by the NSF under grants 2336840 and 2211548, by NASA under ULI grants 80NSSC21M0071 and 80NSSC24M0070, and by ONR under grant N00014-22-1-2703.}
}

\maketitle

\begin{abstract}
Noncooperative multi-agent systems often face coordination challenges due to conflicting preferences among agents. In particular, \fix{when agents act in their own self-interest, they may prefer different choices among multiple feasible outcomes}, leading to suboptimal results or even safety concerns. We propose an algorithm named trading auction for consensus (\texttt{TACo}), a decentralized approach that enables noncooperative agents to reach consensus without communicating directly or disclosing private valuations. \texttt{TACo} facilitates coordination through a structured trading-based auction, where agents iteratively select choices of interest and provably reach an agreement within an \emph{a priori} bounded number of steps. A series of numerical experiments validate that the termination guarantees of \texttt{TACo} hold in practice, and show that \texttt{TACo} achieves a median performance that minimizes the total cost across all agents, while allocating resources significantly more fairly than baseline approaches.
\end{abstract}


\section{Introduction}

\begin{figure}[t]
    \centering
    \includegraphics[width=0.9\linewidth]{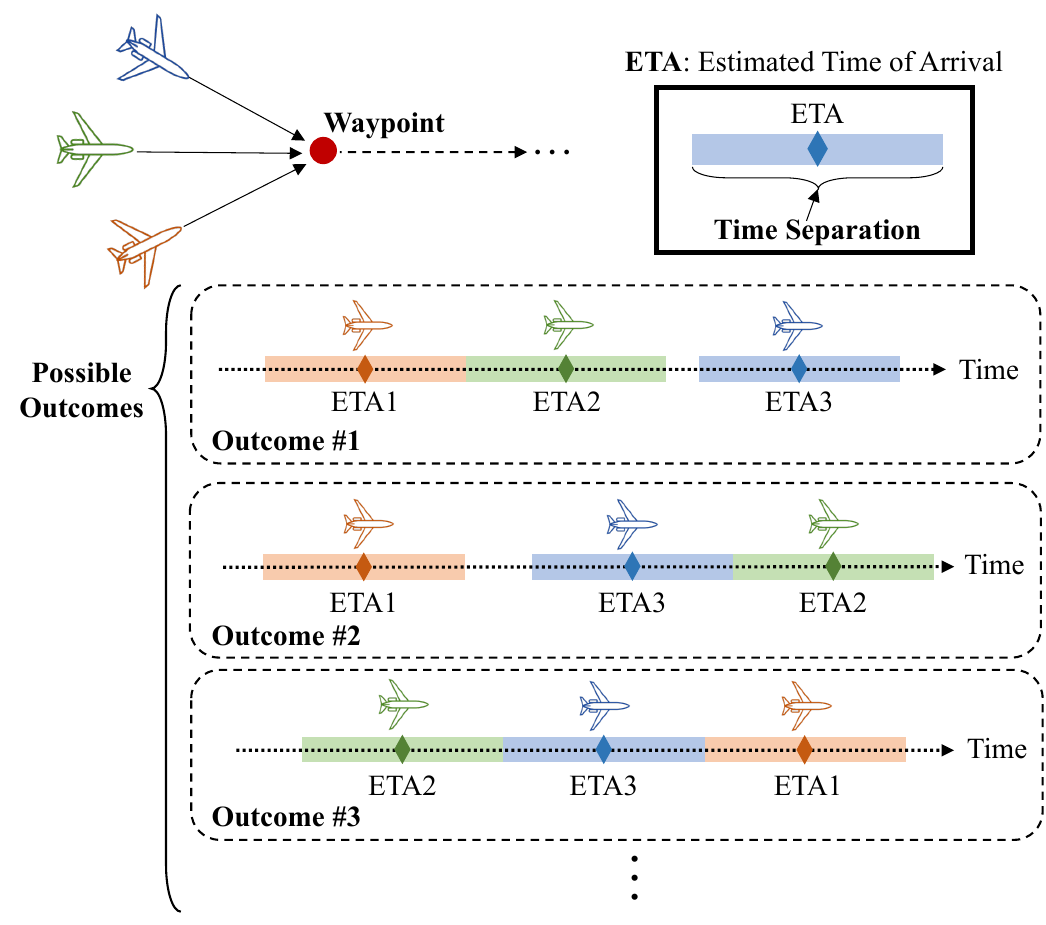}
    \caption{Illustration of a waypoint merging scenario in air traffic management, 
    where multiple aircraft must coordinate their estimated times of arrival (ETAs) at a shared waypoint. 
    \fix{Each possible outcome represents} a feasible ordering of aircraft that maintains safe time separation, 
    highlighting the need for consensus among self-interested agents to avoid conflicts.}
    \label{fig:scenarioDescription}
\end{figure}

Multi-agent systems often require consensus among agents, yet such consensus can be challenging due to conflicts and disagreements. 
These conflicts stem from differences in the strategic priorities of the agents, making coordination difficult without communication or external mediation. 
\fix{A representative example of such challenges arises when agents must select one shared outcome from multiple feasible choices that reflect their individual strategic preferences.} Without a mechanism to enforce a consensus, these disagreements can cause suboptimal outcomes and even safety risks in critical applications \cite{infer_1, infer_2} as illustrated in \Cref{fig:scenarioDescription}.

Reaching a consensus in multi-agent systems is particularly difficult when agents are noncooperative, and this difficulty arises from several factors. 
First, the consensus process must be \textit{procedurally rational}, meaning \fixx{that participating noncooperative agents should be able to make protocol-prescribed decisions that are consistent with their self-interest at each step.}
Second, in practical cases, centralized coordination may be unavailable, necessitating decentralized approaches. Additionally, noncooperative agents often avoid sharing private information and lack access to 1-on-1 communication, further complicating coordination.

Existing consensus methods often rely on centralized coordination \cite{Coop_1_monotone, Coop_2_potential, Coop_3_hetero, Coop_6, Coop_8, Coop_9}, cooperation assumptions \cite{nonCoop_4_linear, nonCoop_5_linear, nonCoop_2_shared, nonCoop_1_coord}, or direct communication \cite{Coop_4_comm_priv, Coop_5_comm_priv, Coop_7_comm_priv}, limiting their applicability for noncooperative scenarios. For example, voting—a common mechanism for achieving agreement—often leads to outcomes that some agents reject \cite{nego_shoham}. Therefore, we need a decentralized approach that achieves agreement over conflicting choices, respects individual preferences, and preserves privacy without requiring direct communication.

We propose an algorithm called the trading auction for consensus (\texttt{TACo}) that enables agents to reach an agreement in noncooperative situations. This algorithm is a structured trading-based auction where agents iteratively select choices based on self-interest. Over multiple rounds, agents attempt to persuade others to adopt their preferred choice by offering trades. When disagreements persist, they refine their negotiation strategies, making increasingly precise adjustments to their offers. The process continues until no agent sees a meaningful benefit in further negotiation, ensuring a stable outcome---a property that is provably satisfied within a finite number of steps.

\texttt{TACo} addresses the core challenges of noncooperative consensus by enabling agents to adjust their preferences based on individual benefit, while ensuring decentralization and privacy preservation. It preserves privacy by not disclosing valuations over choices or trading assets. The algorithm does not require a centralized coordinator and uses broadcast communication—a common method of sharing information in multi-agent systems—which avoids direct communication. This comprehensive design makes \texttt{TACo} applicable to a range of noncooperative multi-agent systems, including autonomous vehicles and coordinating drone operations in cities.

The contributions of this work are threefold: (i) We introduce the trading auction for consensus (\texttt{TACo}) algorithm, which enables consensus in noncooperative settings without requiring a centralized coordinator or direct communication, and while preserving privacy. (ii) We prove that \texttt{TACo} enforces an agreement in finite time, and derive an explicit upper bound on the number of steps required for all agents to reach this mutual agreement. (iii) We validate our theoretical findings through numerical experiments and demonstrate that \texttt{TACo} outperforms existing negotiation methods with respect to social optimality and fairness in a practical equilibrium selection scenario.

\section{Related Work}
Existing multi-agent consensus methods employ leader-follower frameworks \cite{consensus_leader_follow1,consensus_leader_follow2}, finite-time consensus algorithms \cite{consensus_finite1,consensus_finite2}, robust control techniques \cite{consensus_robust1}, auction mechanisms \cite{consensus_auction}, and game-theoretic approaches for noncooperative agents \cite{corrEq}. These have been applied in UAV coordination \cite{consensus_formation1, consensus_formation2, consensus_formation3}, traffic flow optimization \cite{corrEq}, and wireless sensor networks \cite{consensus_formation2}. However, most rely on centralized leaders or high-bandwidth communication, limiting their applicability in fully decentralized, noncooperative settings \cite{consensus_review}.

\fix{A related stream of research considers coordination among agents
when multiple feasible outcomes exist, often framed as \emph{multi-choice problems}.}
\fix{Most of these approaches adopt a centralized or cooperative structure,}
allowing agents to collectively evaluate all possible outcomes and agree on one through a common decision rule
\cite{Coop_1_monotone, Coop_2_potential, Coop_3_hetero, Coop_4_comm_priv, Coop_5_comm_priv, Coop_6, Coop_7_comm_priv, Coop_8, Coop_9}.
\fix{While such frameworks can ensure consistency in the final outcome,
they require agents to disclose private cost information  or require agents to follow decision directives issued by a central authority,
making them impractical in decentralized settings where privacy and autonomy are essential.}

\fix{Several works tackle multi-choice problems in noncooperative systems 
\cite{nonCoop_1_coord, nonCoop_2_shared, nonCoop_3_enforce, nonCoop_4_linear, nonCoop_5_linear, findNash_1, findNash_2}.}
Examples include enforcing unique joint decisions by adjusting agents’ utilities
\cite{nonCoop_3_enforce} or employing structured processes like the linear tracing procedure that rely on common initial beliefs
\cite{nonCoop_4_linear, nonCoop_5_linear}.
\fix{Although these methods maintain individual \emph{rationality}, where
agents have incentives to participate based on self-interest, they still depend on partial coordination or shared information, and thus fall short of achieving fully decentralized consensus.}

Negotiation algorithms provide another approach to decentralized consensus among noncooperative agents \cite{nego_shoham}. Common methods include voting, bargaining, and auction-based mechanisms. Voting is widely used but often fails to ensure \textit{rationality}, meaning outcomes may not be acceptable to all agents \cite{nego_shoham}. Bargaining relies on direct communication and is often problem-specific \cite{nego_meyerson_bargain_auction, nego_noncoopBargain, nego_tournament}, limiting its applicability in general decentralized negotiation environments. In contrast, auctions align agents' incentives with self-interest while preserving \textit{privacy} (i.e., agents need not disclose sensitive private information) \cite{nego_bertsekas, nego_meyerson_bargain_auction, nego_rational_auction, nego_shoham, nego_bertsekas_2009}. However, auctions primarily address resource allocation rather than single-choice problems, highlighting the need for research that bridges this gap by adapting auction-based methods for decentralized, noncooperative \fixx{coordination}. 

\section{Coordination Under Conflicting Preferences}

\fix{In decentralized multi-agent systems, agents are often required to reach agreement on one outcome
from a finite set of feasible alternatives, even when their individual preferences conflict.
We refer to this general setting as a \emph{multi-choice consensus problem}.
Each agent aims to minimize its own cost or maximize its utility,
but the system must converge to a single shared choice to ensure coherent operation.
Such situations arise naturally in scheduling, task allocation, and traffic management,
where multiple admissible joint configurations may exist but only one can be implemented in practice.}

A prominent example of this class of problems is the \emph{equilibrium selection problem}.
\fix{In many noncooperative games, several equilibria coexist, each representing a distinct
yet individually stable outcome for all agents.
Although every equilibrium satisfies the best-response condition,
the coexistence of multiple equilibria introduces ambiguity:
agents must somehow agree on one equilibrium to realize a consistent joint outcome.
This subsection illustrates this issue using standard game-theoretic notation.}

\subsection{Equilibrium selection as a representative multi-choice problem}

We model the interaction between agents as a noncooperative game with 
$n \in \intSet_{\ge0}$ agents. \fix{Let $\xSet = \{x_1, x_2, \ldots, x_n\}$ denote the joint action profile of all agents, 
and $\xSet_{\neg i} = \xSet \setminus x_i$ the actions of all agents except $i$.  
Each agent $i \in [n] \equiv \{1,2,\ldots,n\}$ selects an action 
$x_i$ from a feasible action set $\actionPool_i(\xSet_{\neg i})$}.
Each agent $i$ aims to minimize a cost function 
$c_i(x_i, \xSet_{\neg i})$ that depends on its own action $x_i$ 
and those of other agents.  
The collection of cost functions of all agents is denoted as 
$\cSet = \{c_1, c_2, \ldots, c_n\}$.

Generalized Nash equilibrium is a well-known concept that describes possible outcomes in noncooperative settings. It is a point $\mathbf{x}^\star$ where no agent has an incentive to unilaterally change their action, while adhering to constraints that are dependent on other agents' strategies.

\begin{definition}[Generalized Nash equilibrium] \label{def:GeneralizedNashEquilibrium}
A set of strategies $\xSet^\star = \{x_1^\star, x_2^\star, \ldots, x_n^\star\}$ is a generalized Nash equilibrium if the following holds for each agent $i\in[n]$,
\begin{equation}
    c_i(x_i^\star, \xSet_{\neg i}^\star) \leq c_i(x_i, \xSet_{\neg i}^\star),~\forall x_i \in \actionPool_i(\xSet_{\neg i}^\star).
\end{equation}
\end{definition}

At a generalized Nash equilibrium, each agent’s strategy is optimal given the strategies of all other agents and satisfies the feasibility constraints imposed by the other agents' actions. \fix{Unless otherwise stated, all references to Nash equilibria in the remainder of this paper refer to \emph{generalized Nash equilibria}.}

There exist multiple Nash equilibria in many games, meaning that several action profiles $\xSet^\star$ may satisfy the Nash equilibrium condition. This multiplicity arises when different combinations of actions lead to  optimal outcomes for each agent. The presence of multiple equilibria introduces a coordination challenge known as the \emph{equilibrium selection problem}, where agents must agree on one equilibrium from the set of possible options to ensure efficient and safe interaction.

\begin{example}[Two-agent quadratic game]
\label{ex:quadratic_game}
Consider two agents minimizing
\[
c_i(x_i,x_j)=-(x_i+x_j-1)^2,
\quad \text{s.t. } x_i^2+x_j^2\le1.
\]
The corresponding KKT conditions yield two distinct Nash equilibria,
$(x_1^\star,x_2^\star)=\left(\tfrac{1}{\sqrt2},\tfrac{1}{\sqrt2}\right)$
and
$(x_1^\star,x_2^\star)=\left(-\tfrac{1}{\sqrt2},-\tfrac{1}{\sqrt2}\right)$,
demonstrating that even simple games can admit multiple feasible equilibria.
\end{example}

In decentralized systems, achieving consensus on which feasible outcome to select is nontrivial, especially in noncooperative settings where agents have different preferences over choices. Without coordination, there is a risk that agents may converge to different choices, resulting in inefficiencies or even safety risks. Furthermore, agents may not be able to communicate directly or share their preferences, further complicating the multi-choice consensus problem.

\section{Trading auction for consensus}

The Trading Auction for Consensus (\texttt{TACo}) algorithm is designed to address the challenge of decentralized \emph{conflicting choice} problems among non-cooperative agents. The primary goal of \texttt{TACo} is to help agents reach a consensus or agree on a single choice, even when they have conflicting preferences. The algorithm accomplishes this without requiring direct negotiation or communication between agents, and it preserves each agent's private information during the process.

\subsection{Key elements of \texttt{TACo}}

\texttt{TACo} consists of several key components: a pay matrix, an offer matrix, a cost matrix, a profit matrix, the private valuations over secondary assets, the amount of trading units, and the decrement factor. These components are central to how agents compute their profit and make decisions during \texttt{TACo}. Let $n \in \mathbb{Z}_{\geq0}$ denote the number of agents, and $m \in \mathbb{Z}_{\geq0}$ the number of choices available in the auction. For any matrix $X$, $X_{ij}$ refers to its $(i,j)$ entry.

\begin{enumerate}[\hspace{0pt}1)]
    \item \textbf{Offer matrix ($\offerList$):} The offer matrix $\offerList \in \mathbb{R}^{n \times m}$ represents the units that each agent $i$ can receive \fixx{at termination for each possible final choice}. Specifically, \fixx{at termination,} $\offerList_{ij}$ denotes the number of units that agent $i$ will receive when \fixx{the algorithm selects} choice $j$ \fixx{as the final outcome}, determined by the offers made by other agents.
    
    \item \textbf{Pay matrix ($\payList$):} The pay matrix $\payList \in \mathbb{R}^{n \times m}$ captures the \fixx{payment each agent incurs at termination for each possible final choice}. $\payList_{ij}$ is the number of units that agent $i$ has to pay to others when \fixx{the algorithm selects} choice $j$ \fixx{as the final outcome}.
        
    \item \textbf{Cost matrix ($\costList$):} The cost matrix $\costList \in \mathbb{R}^{n \times m}$ reflects the intrinsic cost associated with each \fixx{possible final} choice. $\costList_{ij}$ represents the \fixx{intrinsic cost incurred} by agent $i$ when \fixx{the algorithm selects} choice $j$ \fixx{as the final outcome}. \fix{Each row of $\costList$ corresponds to the private cost structure of an agent and need not be shared with others, preserving privacy in how each agent evaluates the available options.}
    
    \item \textbf{Private valuation ($\valuation$):} Each agent $i$ assigns a private valuation, $b_i$, to the units of the asset being traded in the \texttt{TACo} algorithm. This valuation, represented as the $i$-th element of the vector $\valuation \in \mathbb{R}_{>0}^n$, reflects how much agent $i$ values each unit of the secondary asset, such as carbon emission credits. The value $b_i$ is unique to each agent and remains private, influencing how the agent perceives the offer and pay quantities when calculating their profit.
    \item \textbf{Amount of trading units ($d$):} The constant $d \in \mathbb{R}_{>0}$ represents the amount of secondary assets traded in each auction step. This public quantity directly impacts the updates of the offer and pay matrices, influencing each agent's profit calculation.
    \item \textbf{Trading unit decrement factor ($\decFactor$):} The decrement factor $\decFactor \in (0,1)$ is a parameter used to reduce the number of trading units when a specified condition is met. When triggered, the amount of trading units $d$ is scaled by the decrement factor:
    \begin{equation}
    d \leftarrow \decFactor d.
    \end{equation}
    This reduction adjusts the scale of assets exchanged in the auction, allowing for a controlled decrease in trading volume over time or under certain conditions, \fix{which---later in \Cref{sec:proof}---plays a key role in ensuring termination.} \fix{Without this refinement, the trading process could continue indefinitely with large discrete updates, failing to settle into a consensus.}
    
    \item \textbf{Profit matrix ($\profitList$):} The profit matrix $\profitList \in \mathbb{R}^{n \times m}$ represents the profit each agent gains \fixx{for each possible final choice under the current offer and pay matrices}. Specifically, $\profitList_{ij}$ denotes the profit \fixx{that would be obtained} by agent $i$ when \fixx{the algorithm selects} choice $j$ \fixx{as the final outcome}. The profit is calculated as follows:
    \begin{equation}
        \profitList = \mathrm{diag}(\valuation) \cdot (\offerList - \payList) - \costList.
    \end{equation}
    This equation accounts for each agent's private valuation $\valuation$, the offers received $\offerList$, the payments made $\payList$, and the intrinsic costs $\costList$, resulting in the total profit for each possible choice. \fix{Here, $b_i$ serves as an agent-specific conversion coefficient that normalizes each agent’s intrinsic cost unit in $\costList$ to the common unit of the secondary asset used in trading.}
\end{enumerate}

\fix{The parameters $d$, $\gamma$, and $\varepsilon$ are public protocol constants shared by all agents. They can be predetermined by an oversight authority or jointly agreed upon before execution,
and contain no private information, preserving the decentralized and privacy-preserving nature of the algorithm.}

\subsection{Auction mechanism} \label{sec:AuctionMechanism}

\texttt{TACo} operates through sequential steps, where agents take turns making decisions. At each step, an agent selects the choice they value most and attempts to persuade others by offering payments. These decisions are based on the matrices $\offerList$, $\payList$, and $\costList$. Before explaining the \texttt{TACo} mechanism, we introduce the following definitions.

\begin{definition}[Agent-profit matrix tuple $(i, \profitList)$]  
The agent-profit matrix tuple $(i, \profitList)$ represents the association of agent index $i \in \mathbb{N}$ with a specific profit matrix $\profitList$ applied in the current step.
\end{definition}

A cycle refers to a sequence of auction processes that returns to a previously encountered profit matrix--playing agent pair. Once the process revisits an existing profit matrix--playing agent pair, the sequence repeats, forming a loop.

\begin{definition}[Cycle, $\cycle$]\label{def:cycle}  
A cycle, $\cycle$, is a finite sequence of agent--profit matrix tuples,
\[
\cycle = \left((i^{k_1}, \profitList^{k_1}), (i^{k_2}, \profitList^{k_2}), \ldots, (i^{k_\ell}, \profitList^{k_\ell})\right),
\]
that returns to a previously encountered tuple, i.e.,
\[
(i^{k_{\ell+1}}, \profitList^{k_{\ell+1}}) = (i^{k_1}, \profitList^{k_1}),
\]
\fix{where the $(\ell+1)$-th tuple coincides with the first, and $\ell$ denotes} the length of the cycle.
\end{definition}

\begin{definition}[Step, $k$]  
A step in the \texttt{TACo} process refers to a single update to the profit matrix $\profitList$, which occurs based on the choice made by one agent during that auction step.
\end{definition}

\begin{algorithm} [hbt!]
\caption{Trading Auction for Consensus (\texttt{TACo})}\label{alg:TACo}
\begin{algorithmic}[1]
\Require $n$: number of agents, $m$: number of choices, $\varepsilon$: termination tolerance
\Require $\costList \in \mathbb{R}^{n \times m}$, $d_0 \in \mathbb{R}_{>0}$, $\decFactor \in (0,1)$

\State Initialize offer matrix $\offerList \in \mathbb{\fix{R}}^{n \times m}$ to $\mathbf{0}_{n \times m}$
\State Initialize pay matrix $\payList \in \mathbb{\fix{R}}^{n \times m}$ to $\mathbf{0}_{n \times m}$
\State \textit{selections} $\gets$ empty list $\in \mathbb{N}^n$
\State \textit{isConverged} $\gets$ False
\State $d \gets d_0$ \Comment{Initialize trading unit}
\State \fixx{$k \gets 0$} \Comment{\fixx{Initialize step counter}}
\State \textit{recordedStates} $\gets$ empty set \Comment{For cycle detection}

\While{not \textit{isConverged}}
    \State \fixx{$k \gets k+1$}
    \State Choose a playing agent $i$ in a \fix{cyclic order}.
    \For{each choice $j = 1$ to $m$}
        \State Compute profit for agent $i$ and choice $j$:
        \State \fixx{$\profitList_{ij}^k \gets b_i (\offerList_{ij}^k - \payList_{ij}^k) - \costList_{ij}$}
    \EndFor
    \State \fixx{$j_{\mathrm{sel}}^k \gets \underset{j \in [m]}{\arg\max}(\profitList_{ij}^k)$} 
    \Comment{\fixx{Best response to step-$k$ profit}}
    \State Update the matrices based on the choice \fixx{$j_{\mathrm{sel}}^k$}:
    \State \quad $\payList_{i\fixx{j_{\mathrm{sel}}^k}} \gets \payList_{i\fixx{j_{\mathrm{sel}}^k}} + nd$
    \State \quad $\offerList_{i\fixx{j_{\mathrm{sel}}^k}} \gets \offerList_{i\fixx{j_{\mathrm{sel}}^k}} + d, \forall i\in [n]$
    \State $\text{\textit{selections}}_i \gets \fixx{j_{\mathrm{sel}}^k}$ \Comment{Record agent $i$'s choice}
    \If{current $((\offerList-\payList), i)$ exists in \textit{recordedStates}} \Comment{Detect cycle}
        \State Reduce trading unit: $d \gets \decFactor \cdot d$
        \State Clear \textit{recordedStates}
        \For{each agent $i\in[n]$}
            \State $\max_j \profitList_{ij} - \min_j \profitList_{ij} < \varepsilon, \, \forall j \in \text{cycle}$ \Comment{Check termination condition}
        \EndFor
        \If{condition is met for all agents}
            \State \textit{isConverged} $\gets$ True
        \EndIf
    \EndIf
    \State Add current $((\offerList-\payList), i)$ to \textit{recordedStates}
\EndWhile
\State \fix{$j^\star \gets \text{mode}(\textit{selections})$} \Comment{\fix{Select the most frequent value in \textit{selections}}}
\For{\fix{each agent $i \in [n]$}}
    \State \fix{$\pi_i \gets \payList_{ij^\star} - \offerList_{ij^\star}$}
    \Comment{\fix{Compute one-time payment at termination}}
\EndFor
\State \fix{\Return $j^\star$, $\{\pi_i\}_{i=1}^n$, $\offerList$, $\payList$}

\end{algorithmic}
\end{algorithm}

The following rules define the \texttt{TACo} process:

\begin{enumerate}[\hspace{1pt}1)]
    \item \textbf{Sequential play:} Agents take turns sequentially in a predefined order. For example, if there are three agents, the order of play would be agent 1 $\rightarrow$ agent 2 $\rightarrow$ agent 3, and then the sequence repeats.
    
    \item \textbf{Profit calculation:} During agent $i$'s turn \fixx{at step $k$}, they select the choice $j$ that maximizes their \fixx{profit value}:
    \begin{equation} \label{eq:profUpdate}
    \profitList_{ij}^{\fixx{k}} = b_i (\offerList_{ij}^{\fixx{k}} - \payList_{ij}^{\fixx{k}}) - \costList_{ij},
    \end{equation}
    where the $i$-th row of each matrix corresponds to the information relevant to agent $i$. \fixx{Here, $\profitList_{ij}^k$ denotes the profit agent $i$ would obtain if choice $j$ were selected as the final outcome under the offer and pay matrices at step $k$.}
    
    \item \textbf{Matrix update:} When an agent $ i $ selects a particular option $ j $, the matrices are updated as follows:
    
    \begin{itemize} \setlength{\itemindent}{-0.8em}
        \item The algorithm updates the payment matrix $\payList$ by incrementing the entry $\payList_{ij}$ by $ nd $, where $ n $ is the total number of agents and $ d $ is the trading unit. This reflects that agent $ i $ \fixx{accumulates a payment obligation of} $ n $ units of the secondary asset for selecting option $ j $:
        \begin{equation} \label{eq:payUpdate}
            \payList_{ij} \gets \payList_{ij} + nd.    
        \end{equation}
        \item The algorithm updates the offer matrix $\offerList$ by incrementing the $ j $-th column by $ d $. This ensures that all agents \fixx{accumulate an additional offered unit} of the secondary asset in their offers for option $ j $:
        \begin{equation} \label{eq:offerUpdate}
            \offerList_{ij} \gets \offerList_{ij} + d, \quad \forall i \in [n].    
        \end{equation}
    \end{itemize}

    \item \label{sec:reductionRule} \textbf{Cycle detection and trading unit reduction:} The trading unit $d$ is reduced by the decrement factor $\decFactor$ when a cycle is detected. A cycle is identified by checking whether the same pair of profit matrix $\profitList$ and active agent at a specific step has occurred in any previous steps. Tracking $\profitList$ is equivalent to tracking $(\offerList - \payList)$, since $b_i$ and $C_{ij}$ are constants based on \cref{eq:profUpdate}. Agents can independently detect cycles by monitoring changes in the publicly observable matrices $\offerList$ and $\payList$. \fix{Because the playing agent index is part of each recorded state, a cycle cannot be detected until every agent has taken at least one turn. This ensures that termination is only possible after a full cyclic pass of agent updates.} When a cycle is detected, the trading unit is updated as $d_{r+1} \gets \decFactor d_r$, where $d_r$ denotes the trading unit after $r$ cycle detections, with the initial trading unit defined as $d_0$. Additionally, the current record of profit matrices is cleared to restart the cycle detection process.
    
    \quad This reduction of the trading unit is a common practice in real bargaining scenarios, \fix{such as price negotiations, where parties gradually reduce their adjustment steps as they move toward an agreement} \cite{kahneman2013prospect}. The relationship governing the trading unit after $r$ cycle detections can be expressed as:
    \begin{equation}\label{eq:dr}
        d_r = d_0 \cdot \decFactor^r.
    \end{equation}
    
    \item \textbf{$\varepsilon$-termination:} The auction continues until consensus is reached. 
    The process terminates when the difference between an agent's maximum and minimum 
    profit across all choices in a cycle falls within a fixed tolerance~$\varepsilon$, 
    \fix{which reflects that the agent is effectively indifferent among the available choices within the detected cycle.}

    \item \fix{\textbf{Transfer settlement:} After the termination criterion is satisfied,
the algorithm fixes the final choice $j^\star$ and settles all transfers \emph{once} for $j^\star$. Each agent’s net transfer in asset units is $\pi_i = \payList_{ij^\star} - \offerList_{ij^\star}$.}
\end{enumerate}

\begin{table}[hbt!]
      \caption{\texttt{TACo} running example}
      \label{tab:locations}
      \centering
      \begin{tabular}{cccccc}\toprule
        \textit{Step} & \textit{Agent, $i$} & $\offerList$ & $\payList$ & $\profitList$ & \textit{Selections} \\ \midrule
        
        1 & 1 & $\begin{bsmallmatrix} 0 & 0 \\ 0 & 0 \end{bsmallmatrix}$ & $\begin{bsmallmatrix} 0 & 0 \\ 0 & 0 \end{bsmallmatrix}$ & $\begin{bsmallmatrix} -10 & -4 \\ -7 & -9 \end{bsmallmatrix}$ & $\begin{bsmallmatrix} \underline{2} & \emptyset \end{bsmallmatrix}$ \\
        
        2 & 2 & $\begin{bsmallmatrix} 0 & 1 \\ 0 & 1 \end{bsmallmatrix}$ & $\begin{bsmallmatrix} 0 & 2 \\ 0 & 0 \end{bsmallmatrix}$ & $\begin{bsmallmatrix} -10 & -4.8 \\ -7 & -7.8 \end{bsmallmatrix}$ & $\begin{bsmallmatrix} 2 & \underline{1} \end{bsmallmatrix}$ \\

        3 & \bf{1} & $\begin{bsmallmatrix} 1 & 1 \\ 1 & 1 \end{bsmallmatrix}$ & $\begin{bsmallmatrix} 0 & 2 \\ 2 & 0 \end{bsmallmatrix}$ & $\begin{bsmallmatrix} \bf{-9.2} & \bf{-4.8} \\ \bf{-8.2} & \bf{-7.8} \end{bsmallmatrix}$ & $\begin{bsmallmatrix} \underline{2} & 1 \end{bsmallmatrix}$ \\

        4 & 2 & $\begin{bsmallmatrix} 1 & 2 \\ 1 & 2 \end{bsmallmatrix}$ & $\begin{bsmallmatrix} 0 & 4 \\ 2 & 0 \end{bsmallmatrix}$ & $\begin{bsmallmatrix} -9.2 & -5.6 \\ -8.2 & -6.6 \end{bsmallmatrix}$ & $\begin{bsmallmatrix} 2 & \underline{2} \end{bsmallmatrix}$ \\

        5 & \bf{1} & $\begin{bsmallmatrix} 1 & 3 \\ 1 & 3 \end{bsmallmatrix}$ & $\begin{bsmallmatrix} 0 & 4 \\ 2 & 2 \end{bsmallmatrix}$ & $\begin{bsmallmatrix} \bf{-9.2} & \bf{-4.8} \\ \bf{-8.2} & \bf{-7.8} \end{bsmallmatrix}$ & $\begin{bsmallmatrix} \underline{2} & 2 \end{bsmallmatrix}$ \\
        
        \bottomrule
      \end{tabular}
\end{table}
\begin{example}[\texttt{TACo} running example] \label{example:2}
    Consider a case with two agents ($n=2$) and two choices ($m=2$). The initial cost matrix is given by $\,\costList=\begin{bsmallmatrix} 10 & 4 \\ 7 & 9 \end{bsmallmatrix}$, and the agents' private valuations for the traded asset are $\valuation=\begin{bsmallmatrix} 0.8 & 1.2 \end{bsmallmatrix}$.
    At each step, agents update their selections based on their calculated profit using the \texttt{TACo} rules. \Cref{tab:locations} shows how the auction progresses over time. At each step, the agent currently playing updates its selection, and this updated selection is \underline{underlined} in the table.
    
    \fix{In this example, agent 1 initially selects option 2 because it provides the highest \fixx{profit value at step 1}. In step 2, agent 2 selects option 1, which provides its highest \fixx{profit value at step 2} given agent 1’s choice. In step 3, agent 1 again chooses option 2 based on the updated profit matrix $\profitList$. In step 4, agent 2 changes its choice and selects option \fixx{2}, and finally, at step 5, agent 1 again selects option 2.}
    
    The \textbf{bold} values in the profit matrix $\profitList$ highlight that the same profit-value and agent pair reappear at steps 3 and 5, signaling the occurrence of a cycle. This cycle is associated with selection 2, where all agents converge to the same option. The algorithm detects that the $\varepsilon$-termination criterion is satisfied because the profit differences for all agents are zero, which is smaller than any positive $\varepsilon$. Consequently, the algorithm terminates at step 5. 
\end{example}

\subsection{Properties of \texttt{TACo}}
There are several properties of the \texttt{TACo} procedure.

\subsubsection{Procedural rationality}
\fixx{In this paper, procedural rationality refers to the rationality of the agents' prescribed actions within the \texttt{TACo} protocol, conditional on participation.}
\texttt{TACo} incentivizes agents to act in their self-interest by enabling them to 
trade and negotiate based on their individual valuations of the available options.  
\fixx{At each step $k$, the playing agent $i$ evaluates the step-wise profit value for each possible final choice $j$ as defined in \Cref{eq:profUpdate}.
This value is the profit that agent $i$ would obtain if choice $j$ were selected as the final outcome under the current offer and pay matrices.}
\fixx{The prescribed update rule is therefore a best response with respect to this step-wise profit: each agent selects the option that maximizes $\profitList_{ij}^k$ among the available choices.}
As demonstrated in \Cref{example:2}, each agent selects the option that maximizes its step-wise profit value $\profitList_{ij}^k$, naturally driving the auction process toward a consensus.
Unlike voting-based methods, which may force majority outcomes without allowing agents to express payoff-based preferences during the coordination process \cite{approvalVoting, nego_shoham}, \texttt{TACo} \fixx{gives each participating agent an opportunity to update its expressed choice according to its own step-wise profit.}

\subsubsection{Privacy preservation}
One of \texttt{TACo}’s core strengths is its ability to preserve the privacy of 
each agent’s valuations. The trading mechanism functions without requiring agents 
to disclose their private valuations for the available options ($\costList$) or 
trading assets ($\valuation$), 
\fix{as \emph{only} the agents’ current choice selections are broadcast at each 
round, and no cost or valuation information is revealed.}
\fix{This contrasts with bargaining-based schemes that rely on pairwise 
exchanges and require disclosure of utility or preference information 
\cite{nego_meyerson_bargain_auction, nego_shoham}.}

\subsubsection{Independence from direct communication}
\texttt{TACo} operates under a minimal communication framework, where agents only 
need to broadcast their choices. This eliminates the need for direct, 1-on-1 
communication, making the algorithm straightforward to implement in practical 
systems.  
\fix{This stands in contrast to bargaining protocols that rely on repeated 
direct exchanges.}

\subsubsection{Termination guarantee}
\texttt{TACo} is guaranteed to terminate with an agreement within a bounded number 
of rounds. The proof of termination, as well as an explicit upper bound on the 
number of rounds needed to terminate, is provided in the following section.  

\subsubsection{\fix{Consensus over discrete choices}}
\fix{\texttt{TACo} enables single discrete-choice consensus among noncooperative agents 
without central coordination. Conventional 
auction mechanisms, by contrast, are typically designed for resource allocation 
\cite{nego_bertsekas, nego_bertsekas_2009, consensus_auction} and do not 
address this problem.}

\begin{remark}[\fixx{Strategic scope of rationality}]
    \fixx{The prescribed update rule is a best response to the step-wise payoff defined by the \texttt{TACo} protocol. Agents are thus modeled as stage-wise rational: each update optimizes the step-wise payoff and does not plan over the future update paths. Forward-looking strategies in which agents infer information from, or strategically influence, the update history are beyond the scope of the present model.}
\end{remark}

\begin{remark} [\fixx{Effect of the order of play}]
    \fixx{As is typical of sequential update dynamics, the order of play shapes the transient evolution through the sequence of updates and the resulting offer and pay matrices; different orderings may therefore result in different final outcomes. However, the transfer mechanism can reduce disparities in realized profit across orderings, although the selected outcome generally remains order-dependent. In repeated applications, randomizing or rotating the initial agent can help avoid systematic first-mover advantages.}
\end{remark}
\section{Termination Proof of \texttt{TACo}} \label{sec:proof}

The proof of \texttt{TACo}'s termination (with an agreement) involves three steps. First, we show that \texttt{TACo} always enters a cycle, where the auction steps repeat in a loop as defined in \Cref{def:cycle}. Next, we demonstrate that profit differences between choices within a cycle are bounded, with the bound proportional to the trading unit per step $d$. This restricts each agent’s available choices to options with profit differences no greater than the bound. Finally, we prove that \texttt{TACo} satisfies the epsilon-termination condition and hence converges. 

The intuition behind the convergence of \texttt{TACo} lies in that, as soon as a cycle is detected, the trading unit $d$ is reduced by a fixed decrement factor $\decFactor$. However, as $d$ decreases, the profit differences within the cycle shrink and must thus eventually fall below the threshold $\varepsilon$ for all agents. At this point, the agents reach a consensus by definition, and \texttt{TACo} terminates.

\begin{figure}[hbt!]
    \centering
\hspace*{-0.01\textwidth}
\resizebox{0.5\textwidth}{!}{
\begin{tikzpicture}[thick, scale=1.1][hbt!]
    \draw[->] (0,0) -- (7.5,0) node[below] {Step$(k)$};
    \draw[->] (0,0) -- (0,3.2) node[left] {$S_i^k$};

    \draw[thin, dotted] (0,2.5) -- (7.3,2.5) node[anchor=east] at (0,2.5) {$S_i^{\max}$};
    \draw[thin, dotted] (0,0.5) -- (7.3,0.5) node[anchor=east] at (0,0.5) {$S_i^{\min}$};
    \draw[thin, dotted] (3,0.5+0.66) -- (4,0.5+0.66);
    \draw[thin, dotted] (4,0.5+0.66*2) -- (5,0.5+0.66*2);

    \node[below left] at (0,0) {$0$};
    \node[below] at (1,0) {$k$};
    \node[below] at (2,0) {$k+1$};
    \node[below] at (3,0) {$k+2$};
    \node[below] at (4,0) {$k+3$};
    \node[below] at (5,0) {$k+4$};
    \node[below] at (6,0) {$k+5$};
    \node[below] at (6.7,-0.07) {$\cdots$};
    \node[below] at (0.5,-0.07) {$\cdots$};

    \draw[thin, dotted] (1,0) -- (1,2.5);
    \draw[thin, dotted] (2,0) -- (2,0.5);
    \draw[thin, dotted] (3,0) -- (3,0.5+0.66);
    \draw[thin, dotted] (4,0) -- (4,0.5+0.66*2);
    \draw[thin, dotted] (5,0) -- (5,2.5);
    \draw[thin, dotted] (6,0) -- (6,0.5);

    \draw[very thick, dashed] (1,2.5) -- (2,0.5);
    \draw[very thick, dashed] (2,0.5) -- (3,0.5+0.66);
    \draw[very thick, dashed] (3,0.5+0.66) -- (4,0.5+0.66*2);
    \draw[very thick, dashed] (4,0.5+0.66*2) -- (5,2.5);
    \draw[very thick, dashed] (5,2.5) -- (6,0.5);
    \node[right] at (6.5,1.5) {$\cdots$};
    \node[left] at (0.82,1.5) {$\cdots$};
    
    \draw[decorate, decoration={brace, mirror, amplitude=8pt}, gray] 
        (0,2.48) -- (0,0.52) node[midway, left=6pt, gray] {$\textstyle d(n-1)b_i$};
    \draw[decorate, decoration={brace, mirror, amplitude=8pt}, gray]
        (3,0.52) -- (3,0.5+0.64) node[midway, right=6pt, gray] {$\textstyle db_i$};
    \draw[decorate, decoration={brace, mirror, amplitude=8pt}, gray]
        (4,0.5+0.66) -- (4,0.5+0.66*2-0.02) node[midway, right=6pt, gray] {$\textstyle db_i$};
    \draw[decorate, decoration={brace, mirror, amplitude=8pt}, gray]
        (5,0.5+0.66*2+0.02) -- (5,2.48) node[midway, right=6pt, gray] {$\textstyle db_i$};
\end{tikzpicture}
}
\caption{(Illustration of \Cref{theorem:lemma1}, $n=4$) This figure \fix{illustrates that} the row sum $S_i^k$ of the $i$-th row of the profit matrix $\profitList$ \fix{returns to the same value every $n$ steps.} The fluctuations in the row sum are bounded by the difference $S_i^{\text{max}} - S_i^{\text{min}} = d(n-1)b_i$, where $S_i^{\text{max}}$ and $S_i^{\text{min}}$ denote the maximum and minimum row sums during the cycle. The diagram highlights the periodic nature of the row sum across steps and the incremental fluctuations $db_i$ within each step.}
\label{fig:fluctuations}
\end{figure}

\subsection{Existence of cycles in \texttt{TACo}}
We first prove that \texttt{TACo} always enters a cycle. A cycle occurs when the auction steps repeat in a loop, as defined in \Cref{def:cycle}. This result is established by analyzing bounded row sums and discrete profit increments. The following lemma shows that the row sum of the profit matrix fluctuates within a bounded range as shown in \Cref{fig:fluctuations}.

\begin{lemma}[Row sum of the profit matrix] \label{theorem:lemma1}
The row sum $S_i^k = \sum_{j=1}^m \profitList_{i,j}^k$, where $m$ is the number of choices, \fix{returns to the same value every $n$ steps}, with fluctuations equal to:
\begin{equation}
    \smax - \smin = d(n-1)b_i,
\end{equation}
where $\smax$ and $\smin$ are the maximum and minimum row sums that can occur during the \texttt{TACo} process.
\end{lemma}

The bounded row sums imply that the actual profit values $\profitList_{ij}$ are lower bounded, as established in the next lemma.

\begin{lemma}[Lower bound of profit values] \label{theorem:lemma2}
The profit value $\profitList_{ij}$ for any agent $i$ and choice $j$ is lower bounded as
\begin{equation}
    \profitList_{ij} \geq \min\left(\min(\profitList_{i,:}^{0}), \textstyle{\frac{\smin}{m}} - d(n-1)b_i\right),
\end{equation}
where $m$ is the number of choices and  $\min(\profitList_{i,:}^{0})$ is the minimum element for row $i$ in the initial profit matrix $\profitList$.
\end{lemma}

Since the row sums of the profit matrix $\profitList$ \fix{return to the same value every $n$ steps} (\Cref{theorem:lemma1}) and since $\profitList$ is lower-bounded (\Cref{theorem:lemma2}), it directly follows that  $\profitList$ is also upper-bounded. We state this consequence in the following corollary.

\begin{corollary}[Upper bound of profit values] \label{theorem:corollary1}
 The profit values in the matrix $\profitList$ are upper bounded.
\end{corollary}

Exploiting the proven upper and lower bounds on $\profitList$ derived so far, we can now formally prove that \texttt{TACo} always enters a cycle. \fix{By definition, consensus corresponds to the condition in which all agents have similar preferences—within a tolerance $\varepsilon$—which itself constitutes a cycle.} Hence, the convergence of \texttt{TACo} will subsequently boil down to showing that the cycle it eventually falls into corresponds to a consensus.

\begin{theorem}[The auction always falls into a cycle] \label{theorem:theorem1}
Let $\mathcal{S}$ be the set of all possible agent-profit matrix tuples $(i, \profitList)$, where $i \in [n]$ and $\profitList$ is a profit matrix satisfying the constraints defined in 
\Cref{theorem:lemma1}, \Cref{theorem:lemma2} and \Cref{theorem:corollary1}. The \texttt{TACo} algorithm must eventually revisit a previously encountered tuple $(i, \profitList) \in \mathcal{S}$, resulting in the formation of a cycle.
\end{theorem}

\subsection{Bounded profit differences between choices in a cycle}

Having established that \texttt{TACo} always enters a cycle, we now focus on bounding the profit differences among the choices within that cycle. More importantly, we will show that the underlying bound in profit differences is \textit{proportional} to $d$; a property that will allow us to prove  \texttt{TACo} eventually converges to a state of agreement between agents, particularly since $d$ is decremented according to \eqref{eq:dr} once a cycle is detected.  To establish this property, we start by introducing the \textit{choice count matrix} (\(\Delta^\cycle\)), which tracks how many times each agent selects each choice during a cycle.

\begin{definition}[Choice count matrix, $\Delta^\cycle$] 
The choice count matrix, denoted as $\Delta^\cycle \in \mathbb{Z}_{\geq 0}^{n \times m}$, represents the number of times each choice is selected by agents during a cycle. Specifically, $\Delta^\cycle_{i,j}$ denotes the number of times agent $i$ selects choice $j$ within the cycle $\cycle$. 
\end{definition}

It follows that the matrix $\Delta^\cycle$ satisfies the following:
\begin{equation} \label{eq:choiceCountToPay}
    \payList^{t+\ell} - \payList^t = nd \cdot \Delta^\cycle,
\end{equation}
\begin{equation} \label{eq:choiceCountToOffer}
    \offerList^{t+\ell} - \offerList^t = d \cdot H_n \Delta^\cycle,
\end{equation}
where $\payList^{t+\ell} - \payList^t$ and $\offerList^{t+\ell} - \offerList^t$ represent the net changes in the pay matrix and offer matrix after each cycle of length $\ell$, respectively, and $H_n = \textbf{1}_n \textbf{1}_n^\top$ is an $n \times n$ matrix with all elements equal to 1, capturing the aggregation of offers received across all agents. Using the definition of the choice count matrix and its relationship shown in \cref{eq:choiceCountToPay} and \cref{eq:choiceCountToOffer}, we establish the structure of $\Delta^\cycle$ which shows that, within a cycle, all agents select each choice the same number of times.

\begin{lemma}[Uniform choice count matrix in a cycle] \label{theorem:lemma3}
The choice count matrix $\Delta^\cycle \in \mathbb{Z}_{\geq 0}^{n \times m}$ for a cycle  has the form
\begin{equation}
\Delta^\cycle = \begin{bmatrix}
    c_1 & c_2 & \ldots & c_m \\ 
    \vdots & \vdots & \ddots & \vdots \\ 
    c_1 & c_2 & \ldots & c_m 
\end{bmatrix},
\end{equation}
where each $ c_j \in \mathbb{Z}_{\geq 0} $ represents the number of times that choice $ j $ is selected by each agent within the cycle.
\end{lemma}

This uniform structure naturally leads to the concept of \emph{active choices}: the choices selected by agents during a cycle.

\begin{definition}[Active choices in a cycle, $A^\cycle$] The active choices in a cycle, denoted $A^\cycle$, are the column indices of the choice count matrix $\Delta^\cycle$ with non-zero values.  In other words, $A^\cycle$ is a set of indices representing the choices selected at least once by some agent within a cycle.
\end{definition}

We now analyze the profit differences within a cycle by focusing on active choices only. Leveraging \Cref{theorem:lemma1}, \Cref{theorem:lemma2}, and \Cref{theorem:lemma3}, we prove a key result that bounds these profit differences \textit{proportionally} to the amount of trading units $d$.

\begin{theorem}[Bound on profit differences within a cycle] \label{theorem:theorem2}
The profit difference among the active choices $A^\cycle$ in a cycle $\cycle$ for agent $i$ is bounded as 
\begin{equation} \label{eq:profDiff_Theorem2}
U_\cycle^i - L_\cycle^i \leq (m+1)d(n-1)b_{\max},\quad \forall i \in [n],
\end{equation}
where $m$ is the total number of choices, $d$ is the trading unit for the current cycle, $b_{\max} = \max_{i \in [n]} (b_i)$ is the highest private valuation among the agents, and $U_\cycle^i$ and $L_\cycle^i$ are the maximum and minimum profit values for the active choices of agent $i$ within cycle $\cycle$, respectively.
\end{theorem}

\subsection{Finite termination property of \texttt{TACo}}
With the profit differences bounded proportionally to $d$ (\Cref{theorem:theorem2}), we now demonstrate that \texttt{TACo} satisfies the epsilon-termination condition, leading to an approximate agreement in a finite number of steps. Recall that each time a cycle is detected -- which provably happens according to Theorem \ref{theorem:theorem1} -- the algorithm reduces the trading unit $d$ by a fixed decrement factor $\decFactor$. This reduction ensures that the profit differences between choices eventually fall below the threshold $\varepsilon$, leading to termination.

\begin{theorem}[Termination of \texttt{TACo}] \label{theorem:theorem3}
The \texttt{TACo} algorithm terminates in a finite number of steps.
\end{theorem}

We also provide a theoretical bound on the worst-case number of steps \fix{that \texttt{TACo} may spend within cycles until termination}. This bound accounts for the number of cycles needed to meet the epsilon-termination condition as well as the steps within each cycle, hence providing practical insights into the efficiency of \texttt{TACo}.

\begin{theorem}[\fix{Post-cycle termination bound for \texttt{TACo}}] \label{theorem:theorem4}
     \fix{The total number of steps that \texttt{TACo} can spend within cycles 
    before termination} is at most:
    \begin{equation} \label{eq:terminationBound}
        \textstyle{
        \left\lceil 
            \log_{\decFactor} \!\left(\frac{\varepsilon}{(m+1)d_0(n-1)b_{\max}} \right) 
        \right\rceil 
        \cdot 
        n \big((m+1)(n-1)\big)^{nm}
        },
    \end{equation}
    with decrement factor $\decFactor$ and initial trading unit $d_0$.
\end{theorem}

\Cref{theorem:theorem4} implies that increasing $\varepsilon$ to
$\varepsilon/\decFactor$ \fix{reduces the worst-case termination bound by
one entire cycle, \ie, it decreases the maximum number of admissible
cycles by one.}
\section{Numerical Experiments}
\subsection{Waypoint merging scenario}

We model a common noncooperative scenario in air traffic management, 
where $n$ aircraft converging on a waypoint must adjust their speeds to avoid conflicts,
as shown in \Cref{fig:scenarioDescription}. 
Each aircraft $i$ in this \emph{waypoint merging scenario} has an estimated time of arrival (ETA) $e_i$
and an urgency factor $k_i>0$ indicating its sensitivity to delay. 
The adjusted arrival time for aircraft $i$ is $u_i = e_i + x_i$, 
where $x_i$ denotes the chosen speed adjustment. 
We introduce carbon emission credits as a tradable secondary asset.

\fix{A central coordinator, such as a flow management unit, computes a set of 
safety-feasible arrival sequences. In our setting, we assume that the coordinator’s 
role is limited to ensuring safe separation and the selection among feasible options is left to the 
stakeholders. Formally, the coordinator minimizes the aggregate delay cost}
\begin{equation} \label{eq:scenarioCost}
\begin{array}{ll}
     \underset{\fix{\{x_i\}_{i=1}^n}}{\mbox{minimize}} & \fix{\sum_{i=1}^{n} k_i x_i^2,} \\
     \mbox{subject to} & |u_i - u_j| \ge \separationDist, \enskip \forall i \neq j.
\end{array}
\end{equation}
\fix{Because the sign of the terms inside the absolute value in \Cref{eq:scenarioCost} changes depending on the arrival order,
multiple feasible solutions can arise—up to $n!$ in total. 
The coordinator enumerates each solution representing a safe and system-feasible configuration, 
providing a set of options 
$\mathcal{X} = \{x^{(1)}, x^{(2)}, \ldots, x^{(n!)}\}$ to the aircraft.}

\fix{Each aircraft privately evaluates these options according to its own urgency, fuel state, 
and valuation of carbon emission credits, resulting in heterogeneous preferences. 
Reaching agreement on a single shared plan therefore becomes a decentralized 
multi-choice consensus problem.}

Another key challenge is the lack of direct one-on-one communication between aircraft. 
Instead, aircraft use the \acrfull{adsb} system, which broadcasts publicly available information such as positions and ETAs ($e_i$). 
However, \acrshort{adsb} does not support direct negotiation, 
\fix{making it infeasible for aircraft to coordinate their preferences directly.}

\fix{In this setting, each aircraft employs the \texttt{TACo} algorithm to reach agreement on one of the feasible options $\mathcal{X}$. 
By trading emission credits and broadcasting preference updates via \acrshort{adsb}, 
aircraft achieve consensus while acting in their own self-interest.}

\subsection{Baseline algorithms}
We evaluated the effectiveness of the \texttt{TACo} algorithm by comparing it with widely used negotiation approaches. While these baseline methods lack key features of \texttt{TACo}, they serve as intuitive benchmarks for multi-agent coordination. The baseline algorithms considered in this experiment are:

\begin{itemize} \setlength{\itemindent}{-0.8em} 
    \item \textbf{Voting}: Each agent votes for their preferred options, and the final decision is determined by majority rule \cite{approvalVoting}. In the event of a tie, the winner is chosen randomly.
    \item \textbf{Random dictator}: A random agent is selected, and all other agents must conform to the preferences of this chosen agent.
    \item \textbf{Utilitarian}: A central coordinator selects the option that minimizes the total costs across all agents, requiring compliance from every agent.
    \item \textbf{Egalitarian}: A central coordinator selects the option that maximizes fairness among agents, and all agents are required to comply with the decision.
\end{itemize}

The voting mechanism offers a simple, decentralized approach but can lead to outcomes that are not unilaterally rational, as agents must accept majority decisions even when misaligned with their preferences. Moreover, the utilitarian and egalitarian methods optimize system-level objectives but impose centralized decisions, making them unsuitable for noncooperative and decentralized settings. Lastly, the random dictator mechanism also relies on centralized selection but ensures fairness over time in repeated interactions by rotating decision-making power among agents.

\subsection{Evaluation method}

We evaluated the performance of \texttt{TACo} across three key metrics: optimality gap, fairness, and convergence steps.

\begin{itemize} \setlength{\itemindent}{-0.8em}
    \item \textbf{Optimality gap}: The optimality gap quantifies the deviation of the total cost incurred by the algorithm from the optimal total cost achieved by a utilitarian solution. It is defined as:
    \begin{equation}\label{eq:OG}
    \textstyle{
        \mathrm{OG}(\cSet, \cSet^\star) = \frac{\sum_{i \in [n]} c_i}{\sum_{i \in [n]} c_i^\star} - 1,
        }
    \end{equation}
    where $c_i^\star \in \cSet^\star$ is the cost of agent $i$ from the utilitarian solution, and $c_i$ is the cost of agent $i$ under the evaluated algorithm.

    \item \textbf{Gini index}: The Gini index \cite{FairnessGuide} measures fairness among agents. A smaller Gini index indicates more equitable outcomes. It is computed as:
    \begin{equation}\label{eq:GI}
    \textstyle{
        \mathrm{GI}(\textbf{c}) = \frac{1}{2n\sum_{i \in [n]} c_i} \sum_{i,j \in [n]} |c_i - c_j|.
        }
    \end{equation}

    \item \textbf{Convergence steps}: The convergence steps metric measures the algorithm's efficiency by counting the number of steps required for termination.

\end{itemize}

To assess the effectiveness of \texttt{TACo}, we conducted a series of Monte Carlo simulations with 1000 trials to ensure statistical significance. Each trial incorporated randomized initial conditions, including diverse weight parameters ($k_i$) and varying initial estimated times of arrival. Agents' private valuations over secondary assets ($b_i$) were randomly assigned within a predefined range to reflect diverse private information. The decrement factor $\decFactor$ was set to 0.9, with the number of aircraft $n$ set to 4 and the number of choices $m$ set to 24.

The simulations were implemented in the Julia programming language. We used the \texttt{ParametricMCPs.jl} package \cite{juliaParametric} and the \texttt{PATH} solver for mixed complementarity problems \cite{juliaPATH} to compute \fix{the solution of} \Cref{eq:scenarioCost} prior to running \texttt{TACo}.
\footnote{Code available at https://github.com/CLeARoboticsLab/TACo}

\begin{figure}[hbt!]
    \centering
    \begin{subfigure}[b]{0.47\textwidth}
        \includegraphics[width=\textwidth, trim=32 0 7 0, clip]{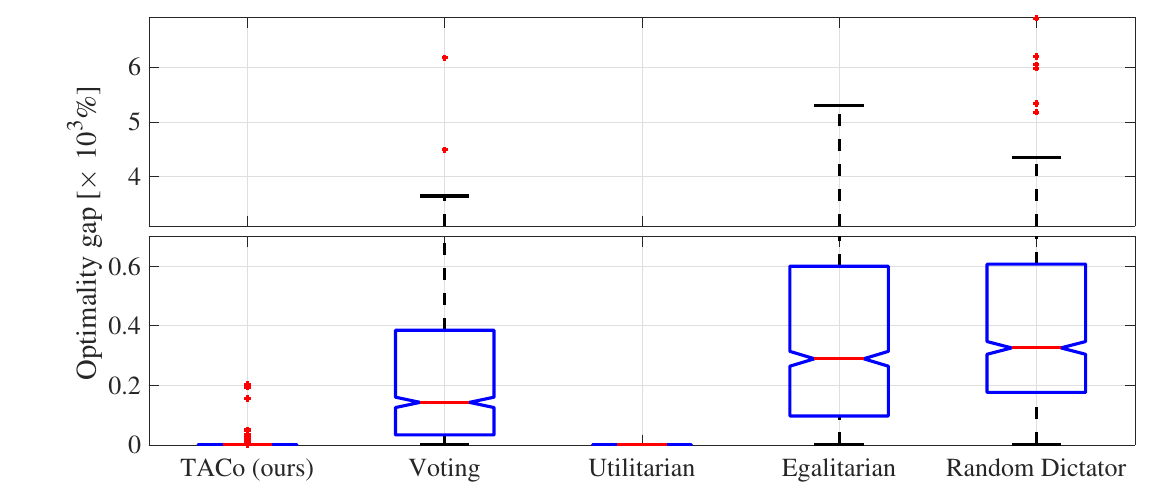}
        \caption{Optimality gap} \label{fig:exp1_optgap}
    \end{subfigure}
    \begin{subfigure}[b]{0.47\textwidth}
        \includegraphics[width=\textwidth, trim=-10 0 0 0, clip]{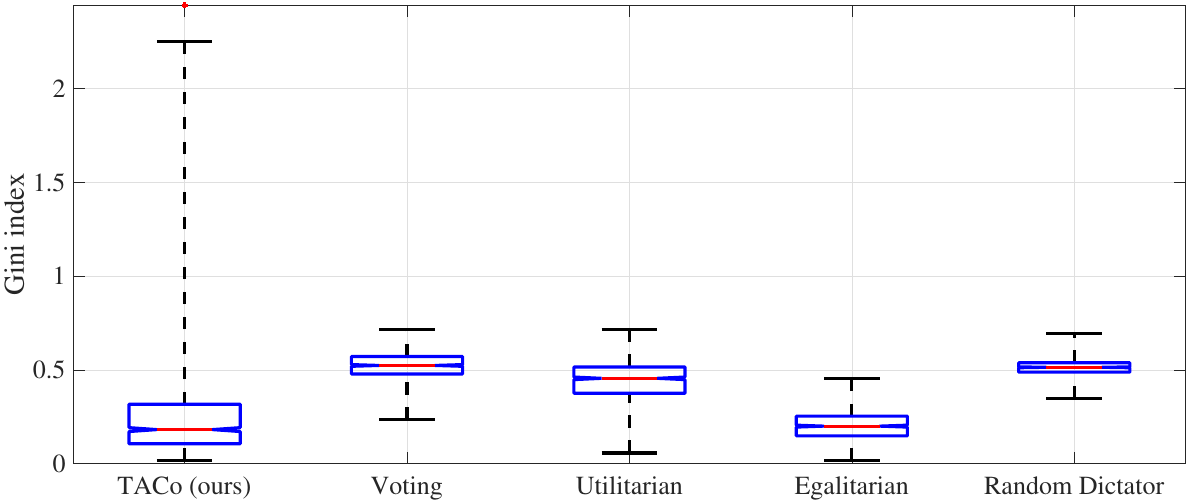}
        \caption{Gini index} \label{fig:exp2_gini}
    \end{subfigure}
    \caption{Experimental results comparing the performance of the \texttt{TACo} algorithm with baseline methods (Voting, Utilitarian, Egalitarian, and Random dictator) across two evaluation metrics: optimality gap and Gini index. \texttt{TACo} demonstrates the smallest median values for the optimality gap and Gini index, highlighting its effectiveness in achieving near-optimal and fair outcomes.}
    \label{fig:result}
\end{figure}

\subsection{Results}

\subsubsection{Optimality and fairness}
As shown in \Cref{fig:exp1_optgap}, the \texttt{TACo} algorithm exhibited a consistently lower optimality gap compared to the baseline algorithms, except for Utilitarian, which achieves a zero optimality gap by definition. \texttt{TACo} showed minimal deviation from the optimal solution, with the median and third quartile value of 0\% and a maximum value of 20.3\%. This \fix{empirically supports} that the \texttt{TACo} effectively minimizes deviations from the socially optimal solution.

\Cref{fig:exp2_gini} shows that \texttt{TACo} achieves a low Gini index, with a median of 0.181, demonstrating that \texttt{TACo} improves fairness by redistributing costs among agents. Notably, trading-based coordination enhances fairness beyond merely selecting the most balanced \fix{choice}. However, in cases where the cost differences between choices were initially small, 
the trading process amplified disparities, resulting in occasional outliers. 
\fix{
This behavior arises when the intrinsic costs~$\costList$ are small relative to the trading unit~$d$ and the termination tolerance~$\varepsilon$, as the discrete trading updates can dominate the underlying cost structure and make outcomes appear less sensitive to the agents’ original cost profiles. 
This observation highlights the importance of selecting~$d$ and~$\varepsilon$ 
at scales consistent with the magnitude of~$\costList$, 
so that the intrinsic cost differences are properly reflected in the consensus outcome.}

\begin{figure}[hbt!]
    \centering
    \includegraphics[width=1.0\linewidth]{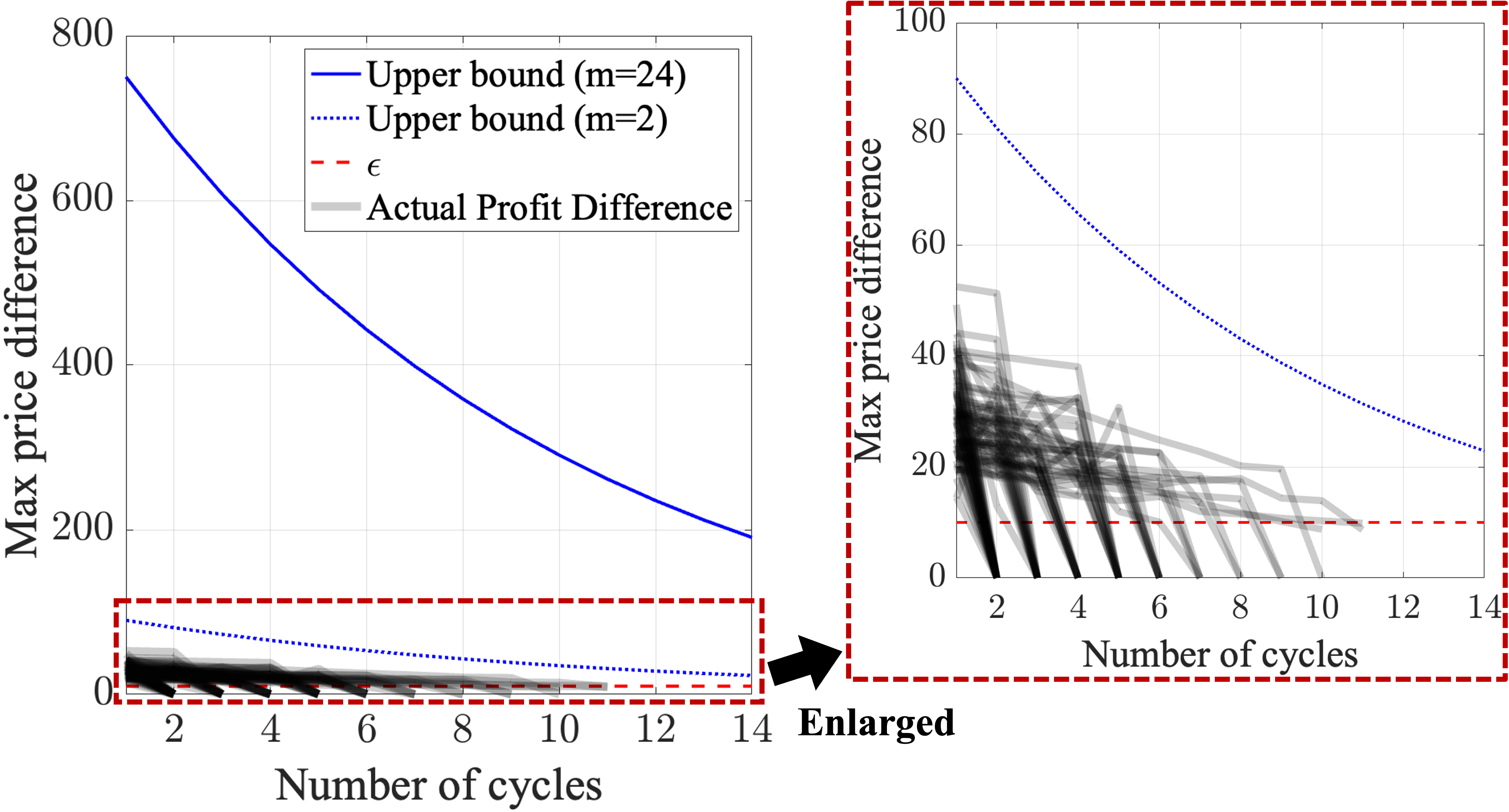}
    \caption{Observed profit differences between choices in cycles, compared with theoretical upper bounds. The solid blue line represents the maximum upper bound derived from \Cref{eq:profDiff_Theorem2} with $m=24$, the maximum number of active choices in this scenario. Since most cycles involved only two choices, a dotted line provides an additional upper bound for $m=2$ for comparison. The results confirm that profit differences consistently remained below the theoretical bounds.}
    \label{fig:price_diff}
\end{figure}

\subsubsection{Termination proof validation}
In our experiments, \texttt{TACo} reached consensus in a median of 53 steps, with a maximum of 380 steps. Given an \acrshort{adsb} information exchange rate of 1 Hz, this corresponds to a median negotiation time of 53 seconds and a maximum of 6 minutes and 20 seconds. While \texttt{TACo} is iterative by design, these results suggest that its convergence remains within operationally feasible limits.

As shown in \Cref{fig:price_diff}, the maximum observed profit difference between choices remained below the theoretical upper bound from \Cref{eq:profDiff_Theorem2}, validating that profit differences decrease as predicted. This empirical confirmation supports the theoretical termination guarantee of \texttt{TACo}.


\begin{figure} [t!]
    \centering
    \begin{subfigure}[b]{0.45\textwidth}
        \includegraphics[width=\textwidth, trim=0 0 -68 0, clip]{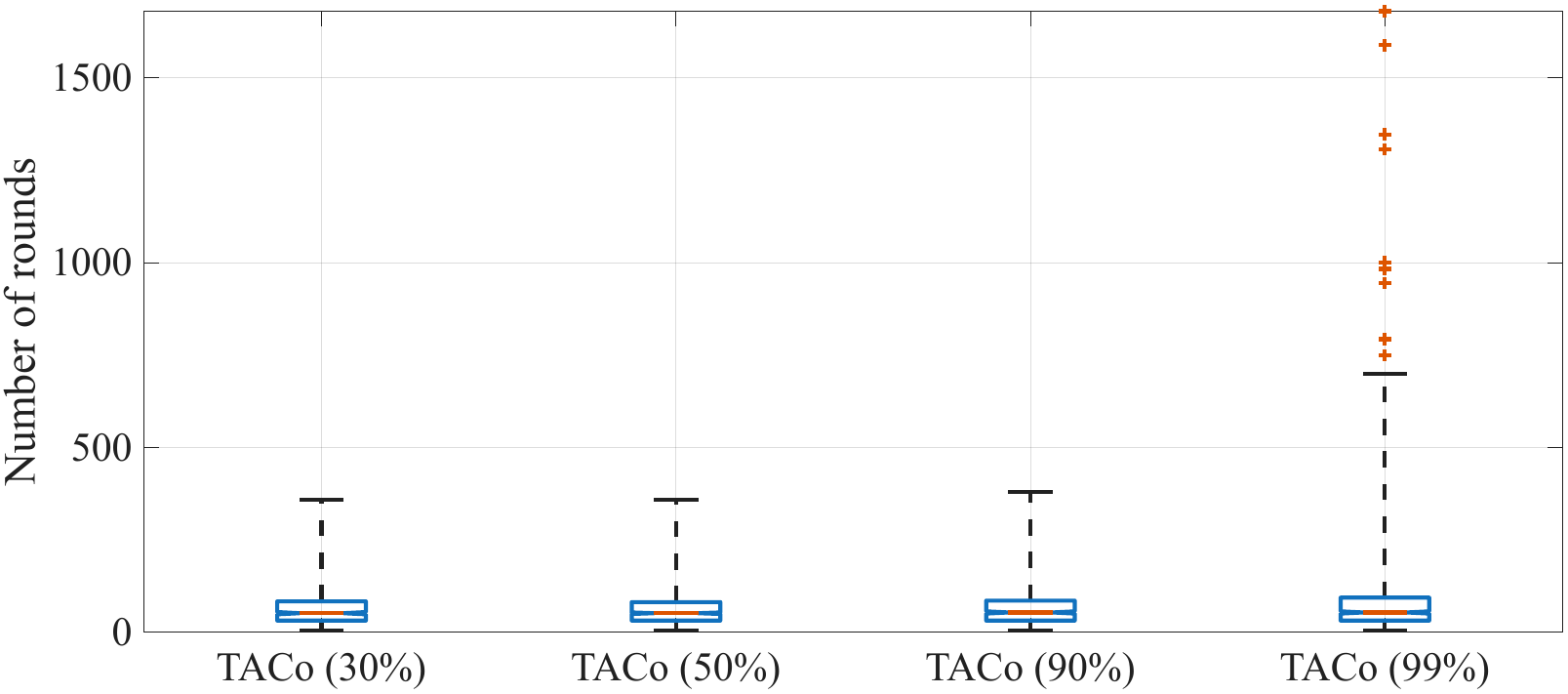}
        \caption{Termination steps} \label{fig:exp4_1_interrupt}
    \end{subfigure}
    \begin{subfigure}[b]{0.45\textwidth}
        \includegraphics[width=\textwidth]{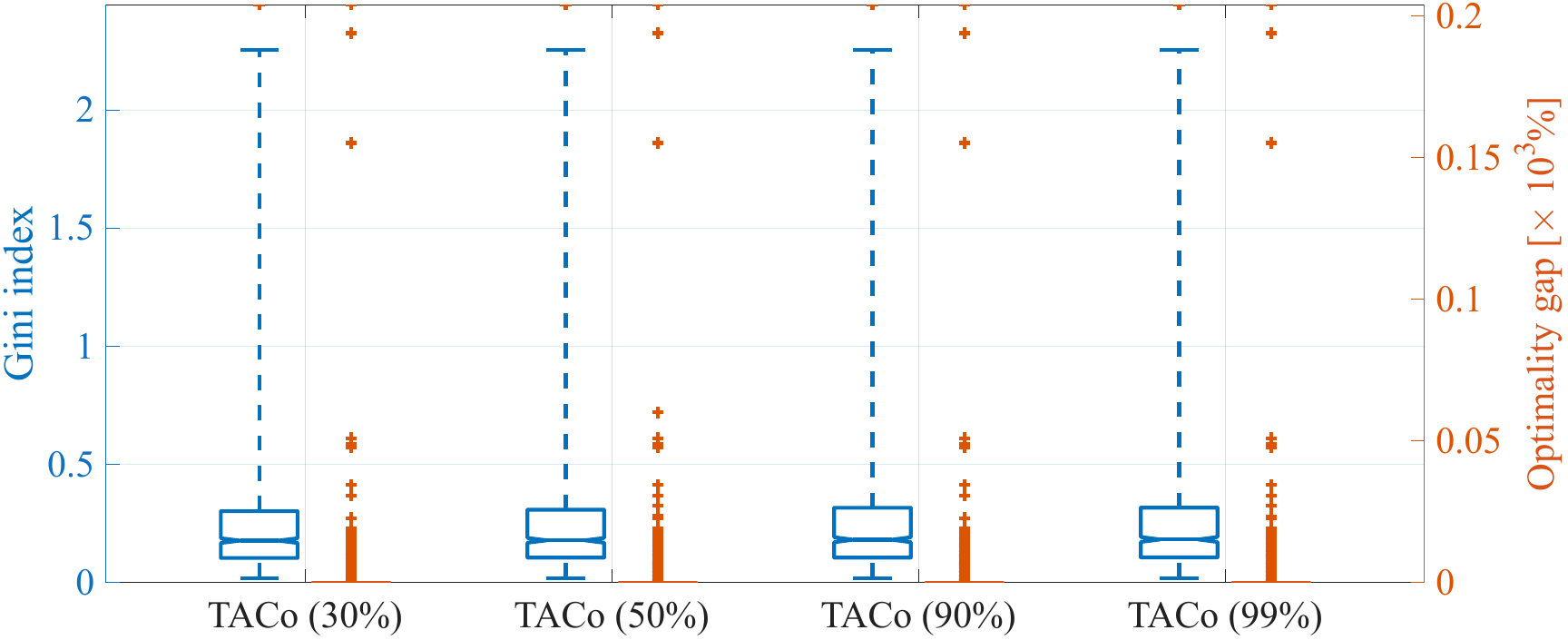}
        \caption{Optimality gap and Gini index} \label{fig:exp4_2_interrupt}
    \end{subfigure}
    \caption{Effect of the decrement factor on \texttt{TACo} performance.  
(a) The termination steps increase as the decrement factor $\decFactor$ approaches 1, with a maximum exceeding 1500 steps. For $\decFactor \leq 0.9$, there is no notable reduction in convergence steps.
(b) The Gini index and the optimality gap remain largely unaffected by changes in $\decFactor$, demonstrating that \texttt{TACo} maintains fairness and optimality regardless of the decrement factor.}
    \label{fig:exp4_tradingStep}
\end{figure}

\begin{figure}[t!]
    \centering
    \begin{subfigure}[b]{0.45\textwidth}
        \includegraphics[width=\textwidth, trim=0 0 -68 0, clip]{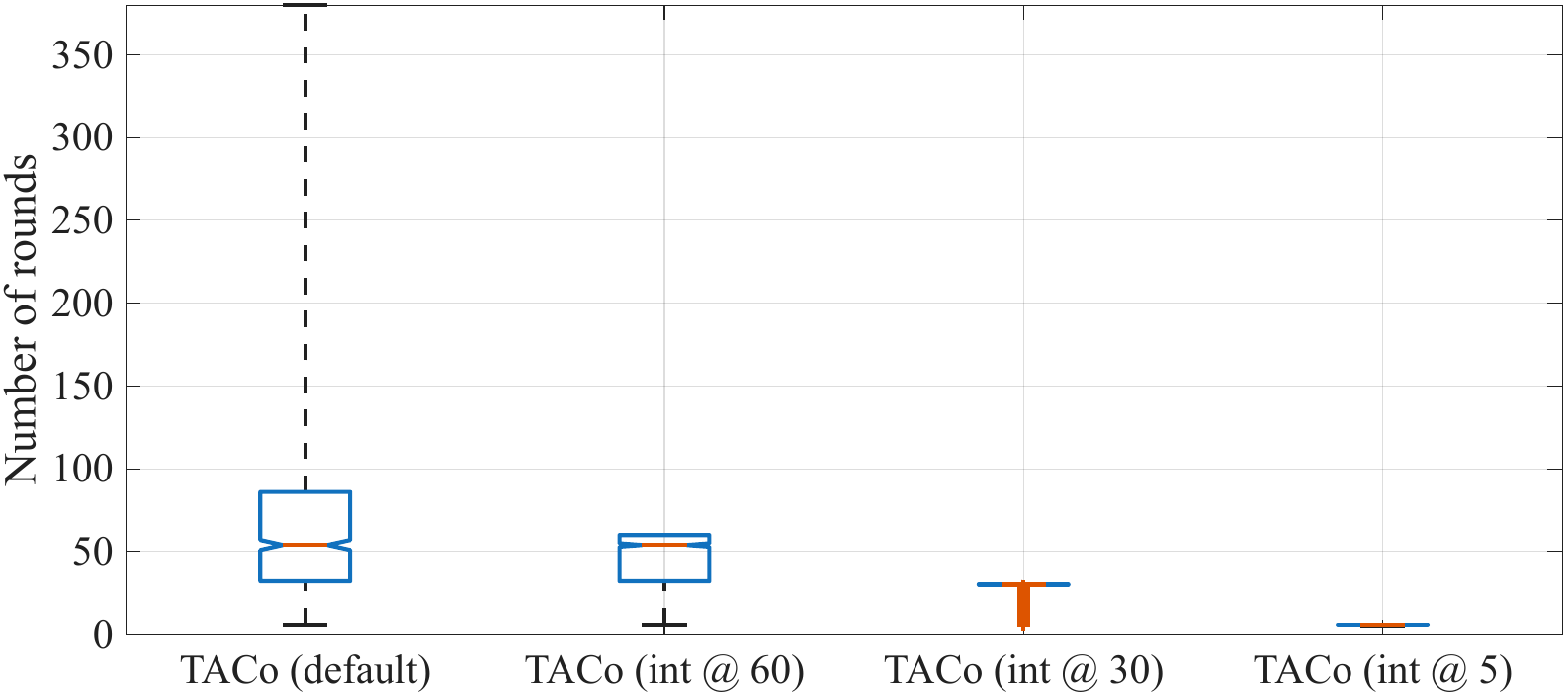}
        \caption{Termination steps} \label{fig:exp5_1_interrupt}
    \end{subfigure}
    \begin{subfigure}[b]{0.44\textwidth}
        \includegraphics[width=\textwidth]{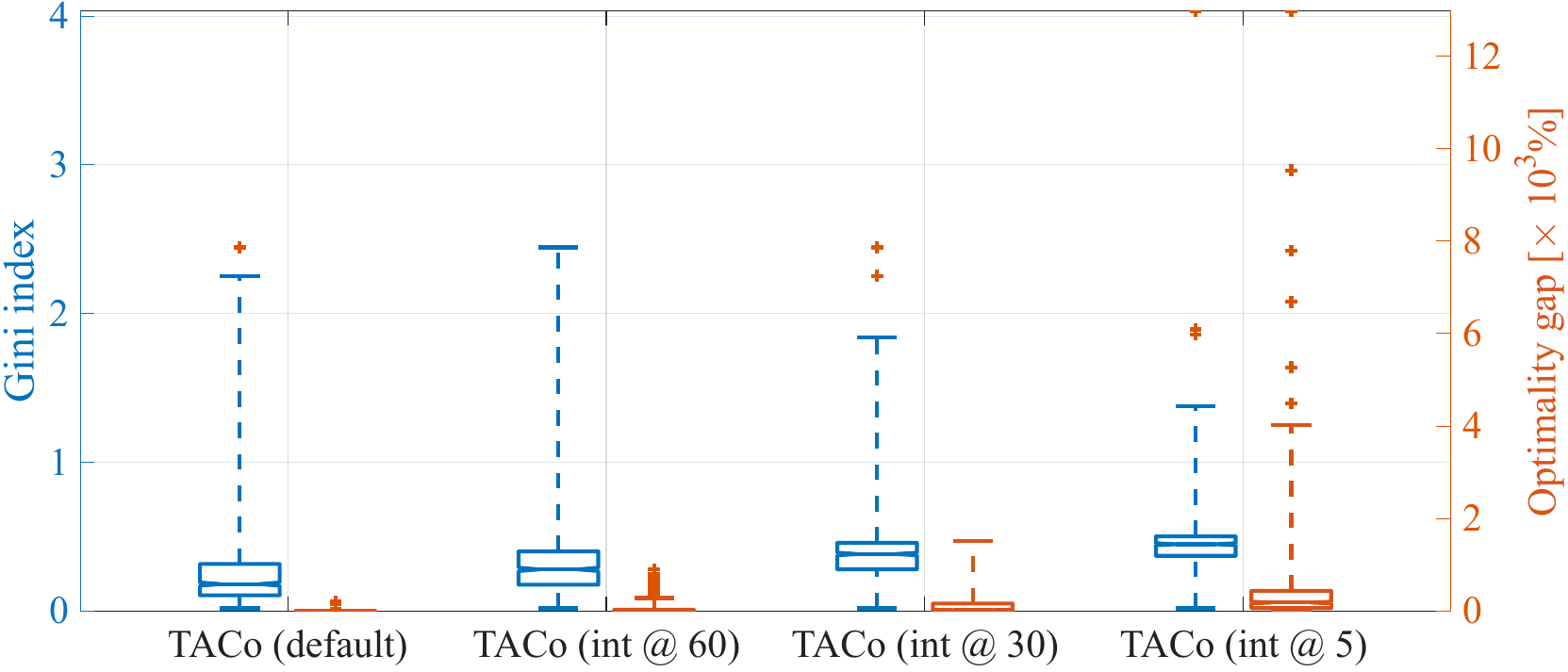}
        \caption{Optimality gap and Gini index} \label{fig:exp5_2_interrupt}
    \end{subfigure}
    \caption{Effect of interruptions on \texttt{TACo} performance.  
(a) Interrupting the algorithm earlier reduces termination steps effectively.
(b) However, earlier interruptions increase the optimality gap and Gini index. This demonstrates the trade-off between faster convergence and reduced fairness and optimality.}
    \label{fig:exp5_interrupt}
\end{figure}

\subsection{\texttt{TACo} variations for faster convergence}

To accelerate the convergence speed of the \texttt{TACo} algorithm, we explored two alternative strategies: (i) adjusting the decrement factor and (ii) interrupting the algorithm to enforce consensus earlier. These experiments highlight how convergence speed can be improved, albeit with potential trade-offs in rationality, optimality, and fairness.

\subsubsection{Effect of decrement factor on convergence}
We investigated how varying the decrement factor $\decFactor$ impacts both convergence steps and performance. In this experiment, we tested $\decFactor$ values ranging from 0.3 to 0.99. As shown in \Cref{fig:exp4_1_interrupt}, the highest number of steps occurred in the $\decFactor=0.99$ case, with a median of 57 and a maximum of 1680 steps. However, for $\decFactor$ values below 0.9, there was no significant improvement in the convergence rate, even for the $\decFactor=0.3$ case.

A similar trend was observed in the Gini index and optimality gap, as shown in \Cref{fig:exp4_2_interrupt}. Just as there was no significant improvement in the convergence rate for smaller $\decFactor$ values, there was also no notable impact on optimality or fairness. Thus, as long as $\decFactor$ does not exceed 0.9, there is no significant impact on convergence steps, optimality, or fairness. 
\fix{ Intuitively, $\gamma$ controls the rate at which the trading unit $d$ shrinks, while $\varepsilon$ determines the final refinement level required for termination. This explains why $\gamma$ affects convergence speed but has limited impact on fairness or optimality in our experiments.
}

\subsubsection{Effect of algorithm interruption on convergence}
In this experiment, we interrupted the algorithm before reaching natural consensus to simulate scenarios where agents must reach a decision within a limited time. Agents were forced to select the most commonly chosen option up to the interrupted step, sacrificing individual rationality for the sake of faster decision-making.

\Cref{fig:exp5_interrupt} shows the results of these interruptions, comparing termination steps and performance across varying levels of interruption. The results demonstrated that interruptions significantly reduced convergence steps, as shown in \Cref{fig:exp5_1_interrupt}. However, as expected, forced consensus increased both the optimality gap and Gini index, as seen in \Cref{fig:exp5_2_interrupt}.

In the most extreme case, where the algorithm was interrupted at the fifth step—allowing each agent only a single chance to express their preference—the median optimality gap increased from 0\% to 182.7\%, and the median Gini index rose from 0.181 to 0.449. This experiment demonstrates that \texttt{TACo} can be adapted to systems requiring faster convergence by allowing interruptions, albeit at the cost of reduced optimality, fairness, and rationality. Such adaptations may be appropriate in high-stakes environments where timely decision-making is more critical than achieving perfect rationality.

\subsection{\fix{Scalability test}}
\fix{
Although \Cref{eq:terminationBound} provides a worst-case upper bound on the number of
rounds required for termination, the bound is conservative.
To evaluate the algorithm’s practical efficiency, we conducted a scalability test by varying both the number of agents ($n$) and the number of choices ($m$).  
We tested $n \in \{3,5,7,10\}$ and $m \in \{3,10,30,100\}$,
using parameters $\gamma = 0.9$, $\varepsilon = 10^{-1}$, and initial trading unit
$d_0 = 1$.  
For each $(n,m)$ pair, we sampled both the intrinsic cost matrix and the private valuations independently from uniform distributions, with $C_{ij},\, b_i \sim \mathrm{Uniform}(0,1)$. Each configuration was run for 100 trials, and we recorded the mean number of rounds to termination with confidence intervals.
}

\begin{figure}[t]
    \centering
    \includegraphics[width=0.90\linewidth]{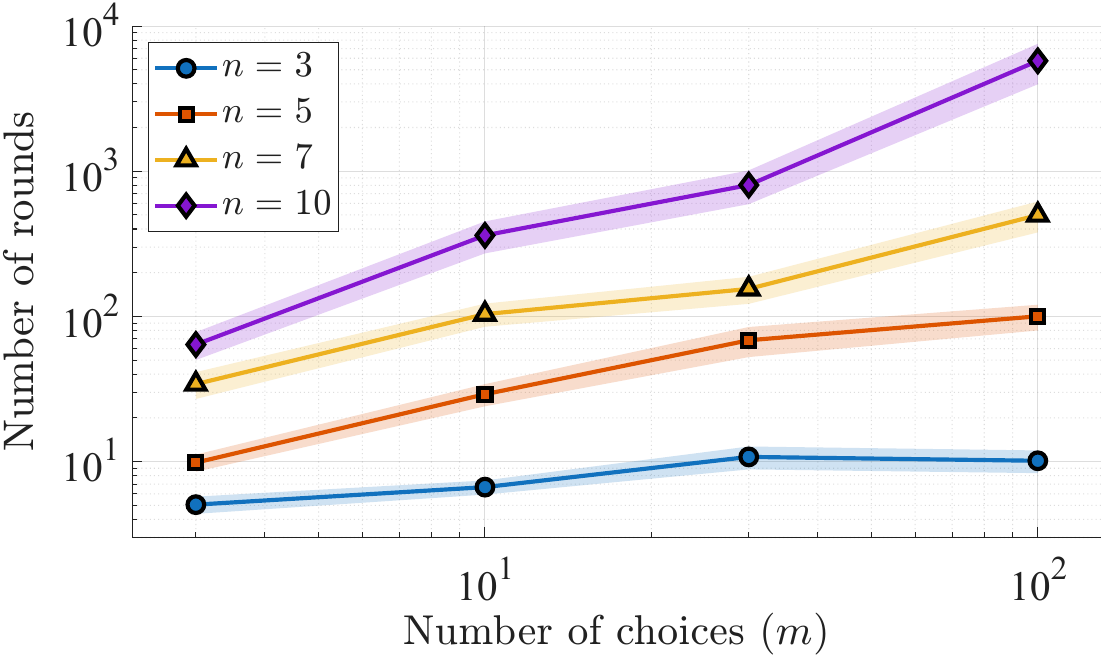}
    \caption{
        \fix{
        Scalability test of \texttt{TACo}, showing the number of rounds required
        until termination.  
        For a fixed number of agents ($n$), the rounds grow sublinearly with respect
        to the number of choices ($m$).  
        For fixed $m$, the rounds increase at a superlinear rate as the number
        of agents ($n$) grows.  
        Overall, the empirical trends indicate that \texttt{TACo} is far more efficient
        in practice than the theoretical worst-case bound in \Cref{eq:terminationBound}.
        }
    }
    \label{fig:complexity}
\end{figure}

\fix{As shown in \Cref{fig:complexity}, the number of rounds grows \emph{sublinearly} with respect to the number of choices~$m$, whereas its dependence on the number of agents~$n$ is significantly stronger, exhibiting growth that exceeds linear scaling for fixed~$m$.
These empirical observations indicate that \texttt{TACo} operates far more efficiently in practice than the theoretical worst-case bound \Cref{eq:terminationBound} suggests, while the bound remains useful for guaranteeing finite-time termination.}

\section{Conclusion}

We developed the trading auction for consensus (\texttt{TACo}) algorithm, a decentralized approach for coordinating decisions among noncooperative agents facing conflicting choices. \texttt{TACo} enables agents to reach consensus by trading secondary assets, ensuring procedural rationality without requiring centralized coordination, direct communication, or the disclosure of private information. 
This flexibility makes \texttt{TACo} an effective tool for resolving coordination challenges in multi-agent systems.

Future work includes improving scalability, evaluating performance in more complex environments, and tightening convergence bounds to ensure \texttt{TACo} terminates quickly in safety-critical settings. This will clarify the types of problems it can efficiently solve and inform strategies for faster convergence.

\section*{References}
\bibliographystyle{IEEEtran}
\bibliography{reference}

\section*{Appendix: Proofs of Main Results}
\begin{customproof}[Proof of \Cref{theorem:lemma1}]
 Over every $n$ steps:
1. During its turn, agent $i$ updates $\payList$ according to \cref{eq:payUpdate}, resulting in a payment of $d(n-1)b_i$ to other agents.
2. During the $n-1$ turns of other agents, agent $i$ receives $db_i$ from each, by updating the offer matrix $\offerList$ according to \cref{eq:offerUpdate}.

Thus, the total amount paid and received for agent $i$ balances out over $n$ steps, keeping the row sum $S_i^k$ constant every $n$ steps. The maximum fluctuation in $S_i^k$ occurs when agent $i$ makes its payment (reducing the row sum by $d(n-1)b_i$), and the row sum is restored during subsequent steps when agent $i$ receives offers. 
\end{customproof}

\begin{customproof}[Proof of \Cref{theorem:lemma2}]
By \Cref{theorem:lemma1}, the row sum $S_i^k$ of the $i$-th row of the profit value matrix is constant every $n$ steps, and the average profit per choice at an arbitrary step $k$ is $\frac{S_i^k}{m}$. In addition, only choices with a profit above this average can be selected by agent $i$, because the choices agent $i$ selects must maximize profit. Moreover, once a choice is selected, its profit decreases by $(n-1)d b_i$. 
As a result, 
 the profit value $\profitList_{ij}$ of choice $j$ for agent $i$ cannot \textit{voluntarily} drop below:
\begin{equation}\label{eq:minprofit}
    \textstyle{\frac{\smin}{m}} - d(n-1)b_i.
\end{equation}
On the other hand, there might exist choices for agent $i$ with a profit value below \eqref{eq:minprofit} due to the way they were initialized. Combining this fact with the \eqref{eq:minprofit}, it follows that 
\begin{equation}
\profitList_{ij} \geq \min\left(\min(\profitList_{i,:}^{0}),\ \textstyle{\frac{\smin}{m}} - d(n-1)b_i\right).
\end{equation}
\end{customproof}

\begin{customproof}[Proof of \Cref{theorem:theorem1}]
By \Cref{theorem:lemma2} and \Cref{theorem:corollary1}, each row of the profit matrix $\profitList$ is bounded above and below, with bounds denoted by $U$ and $L$, respectively. According to Algorithm \ref{alg:TACo}, the profits can vary only in discrete steps of size $d b_i$, where $d > 0$ is the trading unit and $b_i > 0$ is the agent-specific weight. Thus, each element $\profitList_{ij}$ of $\mathcal{J}$ can assume only a finite number of values, given by:
\begin{equation}\label{eq:elementNum}
    \textstyle{\left\lfloor \frac{U - L}{d b_i} \right\rfloor}.
\end{equation}
Since the total number of configurations of $\profitList$ is finite, this means that the set $\mathcal{S}$ of all possible tuples $(i, \profitList)$ is finite as well. As the \texttt{TACo} algorithm iterates, it progresses through elements of this finite set $\mathcal{S}$. By the pigeonhole principle, the \texttt{TACo} algorithm must eventually revisit a previously encountered tuple $(i, \profitList)$ and hence fall into a cycle.
\end{customproof}

\begin{customproof}[Proof of \Cref{theorem:lemma3}]
For a cycle to form, the profit matrix $\mathcal{J}$ must return to a previously encountered state. By \eqref{eq:profUpdate}, this requires that the total amount an agent pays for a specific choice exactly balances the total offers received from other agents for that choice. Mathematically, this condition can be expressed as:
\begin{equation}\label{eq:dPdO}
\payList^{t+\ell} - \payList^t = \offerList^{t+\ell} - \offerList^t.
\end{equation}
From \eqref{eq:choiceCountToPay}-\eqref{eq:choiceCountToOffer}, \eqref{eq:dPdO} is equivalent to
\begin{equation}
nd \cdot \Delta^\cycle = d \cdot H_n \Delta^\cycle,
\end{equation}
from which we obtain
\begin{equation}\label{eq:eig}
(n \textbf{I}_n - H_n) \Delta^\cycle = 0,
\end{equation}
where $\textbf{I}_n$ is the $n \times n$ identity matrix.

Note that in \eqref{eq:eig}, the matrix $(n \textbf{I}_n - H_n)$ is the Laplacian matrix of a fully connected graph with $n$ nodes. As shown in \cite[p.8]{lewis2013cooperative}, for such a Laplacian one must have $\det(n \textbf{I}_n - H_n) = 0$, which indicates the existence of nontrivial solutions to \eqref{eq:eig}. Transforming $(n \textbf{I}_n - H_n)$ into row echelon form using Gaussian elimination yields:
\begin{equation}
n \cdot
\underbrace{
\begin{bmatrix}
    \textbf{I}_{n-1} & -\textbf{1}_{n-1} \\
    \textbf{0}_{n-1}^\top & 0
\end{bmatrix}}_{A} \Delta^\cycle = 0.
\end{equation}
From this structure, the first row of $A$ implies that the first and last elements of each column of $ \Delta^\cycle $ are identical. The second row of $A$ implies that the second and last elements of each column of $ \Delta^\cycle $ are also identical. By induction, all elements within each column of $ \Delta^\cycle $ are identical. Therefore,  $ \Delta^\cycle $ must have the form:
\begin{equation}
\Delta^\cycle = \begin{bsmallmatrix}
    c_1 & c_2 & \ldots & c_m \\ 
    \vdots & \vdots & \ddots & \vdots \\ 
    c_1 & c_2 & \ldots & c_m 
\end{bsmallmatrix},
\end{equation}
where $ c_j \in \mathbb{Z}_{\geq 0} $ for all $ j \in [m] $. This structure implies that, for a cycle to form, each agent must select each choice the same number of times as every other agent within the cycle. This completes the proof.
\end{customproof}

\begin{customproof}[Proof of \Cref{theorem:theorem2}]
From Theorem \ref{theorem:theorem1}, \texttt{TACo} must eventually fall into a cycle $\sigma$. 
Let the number of active choices within that cycle be $p = |A^\cycle|$, and define $C_i$ as the sum of the profits for row $i$ corresponding to the inactive choices, i.e., the choices not in the cycle $\sigma$. Within $\cycle$, the sum of profits for the active choices in row $i$ is:
\begin{equation}
S_i^k - C_i,
\end{equation}
where $C_i$ remains constant during the cycle. Following the same logic as in \Cref{theorem:lemma2}, the lower bound for the profits of the active choices in $\cycle$ becomes:
\begin{equation}
\profitList_{ij} \geq \min\left(\min(\profitList_{i,:}^{\text{0}}), \textstyle{\frac{\smin - C_i}{p}} - d(n-1)b_i\right), \quad \forall j \in A^\cycle.
\end{equation}

From \Cref{theorem:lemma3}, each active choice $j \in A^\cycle$ is selected at least once by each agent. For such a choice to be selected, it must,  at the time of selection, have a profit value greater than the average profit $\frac{\smin - C_i}{p}$ among the active choices in $\cycle$. Accordingly, no choice that has a profit value below $\frac{\smin - C_i}{p}$ may be selected by agent $i$. Therefore, strictly within the active choices in $\sigma$, we obtain the following tighter lower bound on $\mathcal{J}$:
\begin{equation} \label{eq:L_cycle}
\profitList_{ij} \geq \textstyle{\frac{\smin - C_i}{p}} - d(n-1)b_i = L_\cycle^i, \quad \forall j \in A^\cycle.
\end{equation}

Since the row sum of the profit matrix is constant over every $n$ steps, this lower bound ensures the existence of an upper bound for the profits of the active choices. The upper bound, $U_\cycle^i$, occurs when one active choice achieves a profit value above the average, while the remaining $p-1$ active choices are at their lower bound. This situation leads to the following relation:
\begin{equation}
L_\cycle^i (p-1) + U_\cycle^i = \smax - C_i.
\end{equation}
Rearranging for $U_\cycle^i-L_\cycle^i$ yields
\begin{equation}\label{eq:tempdiff}
U_\cycle^i - L_\cycle^i = \smax - C_i  - L_\cycle^i p.
\end{equation}
Substituting $L_\cycle^i$  from \cref{eq:L_cycle} in \eqref{eq:tempdiff},
\begin{align}
U_\cycle^i - L_\cycle^i &= \textstyle{\smax - C_i - p\left(\frac{\smin - C_i}{p} - d(n-1)b_i \right)} \\
&= \smax - \smin + p d(n-1)b_i.
\end{align}

From \Cref{theorem:lemma1}, the row sum difference $\smax - \smin$ is equal to $d(n-1)b_i$. Substituting this result:
\begin{equation} \label{eq:priceDiff}
U_\cycle^i - L_\cycle^i = d(n-1)b_i + p d(n-1)b_i = (p+1)d(n-1)b_i.
\end{equation}

Since $p \leq m$, where $m$ is the total number of choices, the profit difference is bounded as:
\begin{equation} \label{eq:priceDiffUpper}
U_\cycle^i - L_\cycle^i \leq (m+1)d(n-1)b_i \leq (m+1)d(n-1)b_{max},
\end{equation}
for all $i\in[n]$. This concludes the proof.
\end{customproof}

\begin{customproof}[Proof of \Cref{theorem:theorem3}]
By \Cref{theorem:theorem1}, the \texttt{TACo} process always enters a cycle.  
If the cycle involves a single active choice, no profit differences arise and the
$\varepsilon$-termination condition is automatically satisfied.

For cycles involving multiple choices, \Cref{theorem:theorem2} implies that, for all
$i\in[n]$,
\begin{equation} \label{eq:profitBound}
\max_{(i,\profitList^k),(i,\profitList^{k'})\in\cycle} 
|\profitList^k_{i,j} - \profitList^{k'}_{i,j'}|
\le U_\cycle^i - L_\cycle^i
\le (m+1)d(n-1)b_{\max},
\end{equation}
for all active choices $j,j'\in A^\cycle$. 
Each cycle detection triggers the reduction rule in \Cref{sec:reductionRule}, so after $r$ detections the trading unit satisfies $d_r=d_0\decFactor^r$, where $\decFactor\in(0,1)$. Hence, \cref{eq:profitBound} gives
\begin{equation} \label{eq:maxProfDiff_reduction}
(m+1)d_r(n-1)b_{\max}
= (m+1)d_0(n-1)b_{\max}\decFactor^r.
\end{equation}
Since $\decFactor<1$, the right-hand side of
\cref{eq:maxProfDiff_reduction} converges to zero. Therefore, for any
$\varepsilon>0$, there exists a finite $R$ such that, for all $r\ge R$,
\begin{equation} \label{eq:maxProfDiff_epsilon}
\max_{(i,\profitList^k),(i,\profitList^{k'})\in\cycle}
|\profitList^k_{i,j} - \profitList^{k'}_{i,j'}|
\le \varepsilon,
\quad \forall i\in[n],\ \forall j,j'\in A^\cycle .
\end{equation}
At this point, all profit differences among active choices in the detected cycle are within $\varepsilon$ for every agent, so the $\varepsilon$-termination criterion is satisfied and the \texttt{TACo} process terminates.
\end{customproof}

\begin{customproof}[Proof of \Cref{theorem:theorem4}]
    The profit difference among the active choices for agent $i$ is given in \Cref{eq:priceDiff}. Substituting this into \Cref{eq:elementNum}, we infer that  
    the number of possible values that $\profitList_{i,j}$ can take in a cycle is at most $(p+1)(n-1)$, where $p$ is the number of active choices.
    
    Since $p \leq m$ and the number of elements in $\profitList$ is $nm$, the maximum number of distinct $\profitList$ configurations is $\big((m+1)(n-1)\big)^{nm}$. Consequently, the maximum number of unique agent-profit matrix tuples $(i, \profitList)$ in a cycle is $n \cdot \big((m+1)(n-1)\big)^{nm}$. This provides an upper bound on the number of steps required to form a cycle, as established in \Cref{theorem:theorem1}.

    The upper bound on the number of cycles $r$ required to satisfy the termination condition is derived from \Cref{eq:maxProfDiff_reduction} and \Cref{eq:maxProfDiff_epsilon}:
    \begin{subequations}
        \begin{align}
            &(m+1)d_0(n-1)b_{\max} \cdot \decFactor^r \leq \varepsilon, \\
            &\textstyle{\implies r = \left\lceil \log_{\decFactor} \left(\frac{\varepsilon}{(m+1)d_0(n-1)b_{\max}}\right) \right\rceil,}
        \end{align}
    \end{subequations}
    with decrement factor $\decFactor$ and initial trading unit $d_0$.

    Combining the number of cycles needed and the maximum steps per cycle, we obtain the upper bound on the number of steps needed for \texttt{TACo} to terminate:
    \begin{equation}
    \textstyle{
        \left\lceil \log_{\decFactor} \left(\frac{\varepsilon}{(m+1)d_0(n-1)b_{\max}} \right) \right\rceil \cdot n \cdot \big((m+1)(n-1)\big)^{nm}.}
    \end{equation}
\end{customproof}

\begin{IEEEbiography}[{\includegraphics[width=1in,height=1.25in,clip,keepaspectratio]{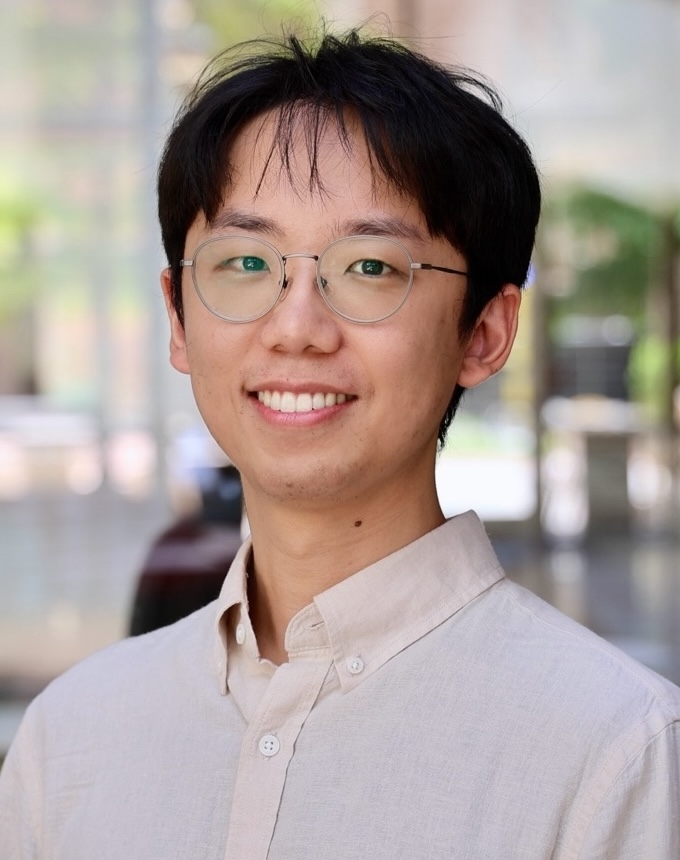}}]{Jaehan Im} (Graduate Student Member, IEEE) received the B.S. and M.S. degrees in Aerospace Engineering from the Korea Advanced Institute of Science and Technology (KAIST), Daejeon, Republic of Korea. He is working toward the Ph.D. degree in Aerospace Engineering and Engineering Mechanics at The University of Texas at Austin, Austin, TX, USA. His research focuses on game-theoretic coordination, robust decision-making, and air traffic management for advanced air mobility.
\end{IEEEbiography}


\begin{IEEEbiography}[{\includegraphics[width=1in,height=1.25in,clip,keepaspectratio]{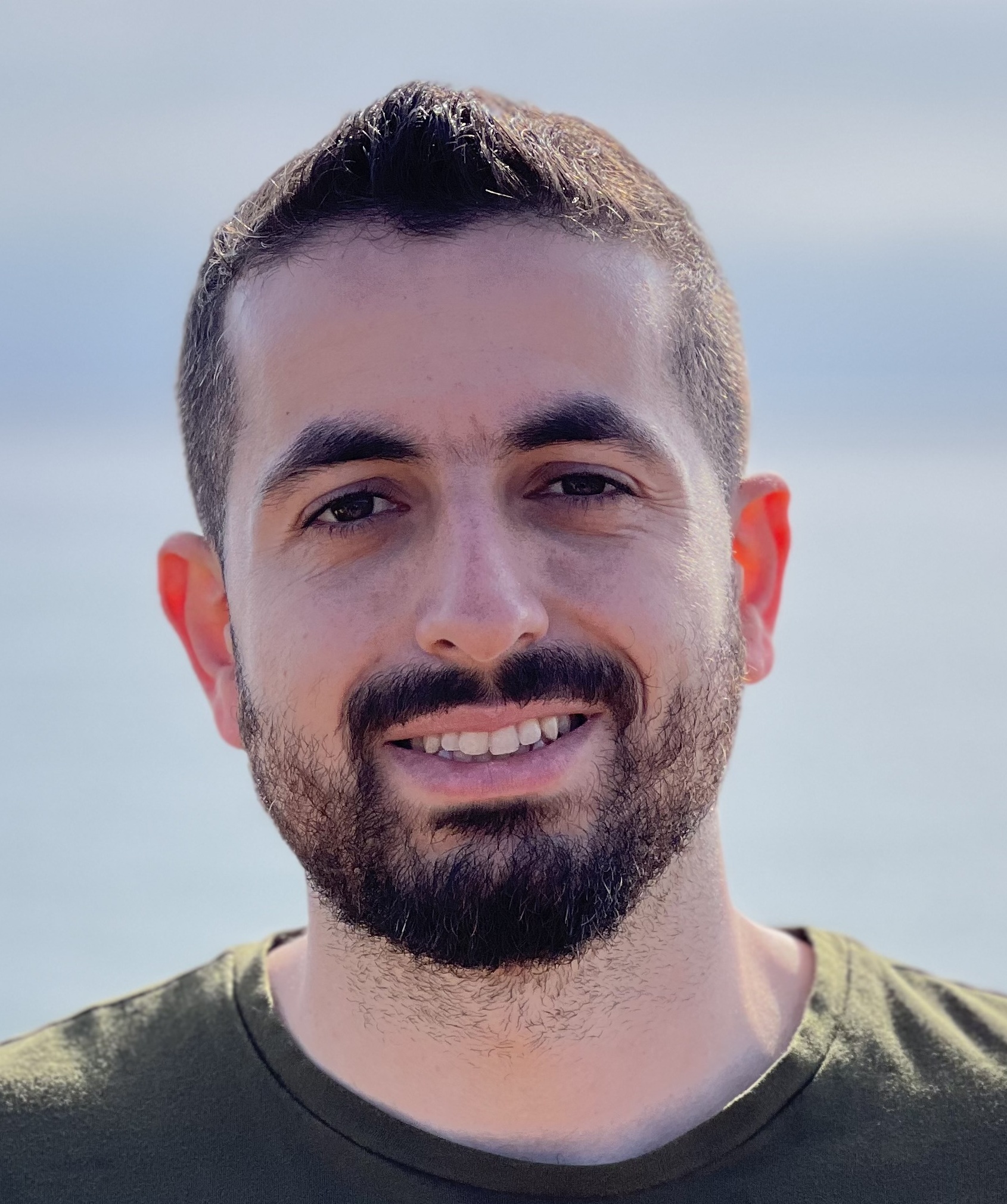}}]{Filippos Fotiadis}
(Member, IEEE) was born in Thessaloniki, Greece. He received the Ph.D. degree in Aerospace Engineering in 2024, and the M.S. degrees in Aerospace Engineering and Mathematics in 2022 and 2023, all from Georgia Tech. Prior to his graduate studies, he received a diploma in Electrical \& Computer Engineering from the Aristotle University of Thessaloniki. He is currently a postdoctoral researcher at the Oden Institute for Computational Engineering \& Sciences at the University of Texas at Austin.
His research interests are at the intersection of systems \& control theory, game theory, and learning, with applications to the security and resilience of cyber-physical systems.
\end{IEEEbiography}


\begin{IEEEbiography}[{\includegraphics[width=1in,height=1.25in,clip,keepaspectratio]{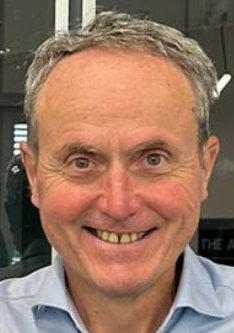}}]{Daniel Delahaye} (Member, IEEE) is the head of the Optimization and Machine Learning Team of the ENAC research laboratory and he is also in charge of the research program “AI4DECARBO” in the new AI institute ANITI in Toulouse and member of the SESAR scientific committee.
He obtained his engineering degree from the ENAC school and did a master of science in signal processing from the National Polytechnic Institute of Toulouse in 1991. He received his Ph.D. in automatic control from the Aeronautic and Space National School in 1995 under the co-supervision of Marc Schoenauer (CMAPX). He did a post-doc at the Department of Aeronautics and Astronautics at MIT in 1996 under the supervision of Pr Amedeo Odoni. He started his career working at the French Civil Aviation Study Center (CENA) and moved to ENAC in 2008. He got his tenure in applied mathematics in 2012. He conducts research on mathematical optimization and artificial intelligence for airspace design and aircraft trajectory optimization.
\end{IEEEbiography}


\begin{IEEEbiography}[{\includegraphics[width=1in,height=1.25in,clip,keepaspectratio]{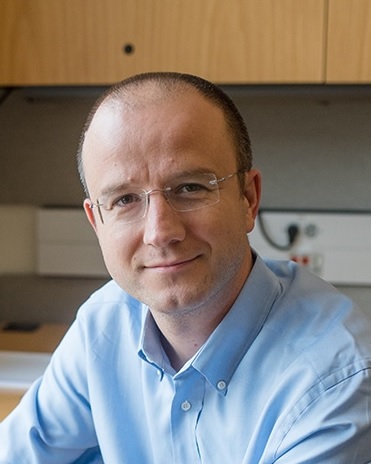}}]{Ufuk Topcu}(Fellow, IEEE)  is currently a Professor
with the Department of Aerospace Engineering and
Engineering Mechanics, The University of Texas at
Austin, Austin, TX, USA, where he holds the Temple
Foundation Endowed Professorship No. 1 Professorship. He is a core Faculty Member with the Oden Institute for Computational Engineering and Sciences and
Texas Robotics and the director of the Autonomous
Systems Group. His research interests include the
theoretical and algorithmic aspects of the design and
verification of autonomous systems, typically in the
intersection of formal methods, reinforcement learning, and control theory.
\end{IEEEbiography}


\begin{IEEEbiography}[{\includegraphics[width=1in,height=1.25in,clip,keepaspectratio]{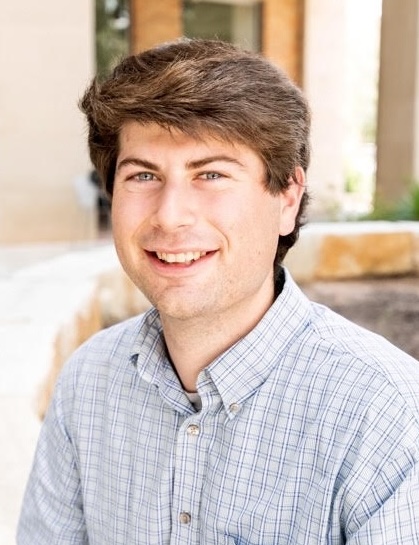}}]{David Fridovich-Keil} (Senior Member, IEEE) received the B.S.E. degree in electrical engineering from Princeton University, and the Ph.D. degree from the University of California, Berkeley. He is an Assistant Professor in the Department of Aerospace Engineering and Engineering Mechanics at the University of Texas at Austin. Fridovich-Keil is the recipient of an NSF Graduate Research Fellowship and an NSF CAREER Award.
\end{IEEEbiography}

\end{document}